\documentclass[twocolumn,twocolappendix]{aastex631}

\usepackage{graphicx}
\usepackage{gensymb}
\usepackage{float}
\usepackage{subfigure}
\usepackage{verbatim}
\usepackage{makecell}
\usepackage{multirow}
\usepackage{appendix}

\begin{document}

\title{Identification of a Radio Counterpart to SN\,2025ulz in the S250818k Localization Area}

\author[0009-0007-1842-7028]{Tanner O'Dwyer}
\affiliation{William H. Miller III Department of Physics and Astronomy, Johns Hopkins University, Baltimore, Maryland 21218, USA.}
\author[0000-0001-8104-3536]{Alessandra~Corsi}
\affiliation{William H. Miller III Department of Physics and Astronomy, Johns Hopkins University, Baltimore, Maryland 21218, USA.}
\correspondingauthor{acorsi2@jh.edu}
\author[0009-0004-7280-1357]{Deepika~Yadav}
\affiliation{Indian Institute of Technology Kanpur, Kanpur 208016, UP, India.}
\author[0000-0002-2557-5180]{Kunal~P.~Mooley}
\affiliation{Indian Institute of Technology Kanpur, Kanpur 208016, UP, India.}
\author[0009-0004-7268-7283]{Raphael~Baer-Way}
\affiliation{National Radio Astronomy Observatory, 520 Edgemont Road, Charlottesville, VA 22903, USA.}
\affiliation{Department of Astronomy, University of Virginia, Charlottesville, VA 22904-4325, USA}
\author[0000-0002-0844-6563]{Poonam~Chandra}
\affiliation{National Radio Astronomy Observatory, 520 Edgemont Road, Charlottesville, VA 22903, USA.}
\author[0000-0002-7083-4049]{Gregg~Hallinan}
\affiliation{Cahill Center for Astronomy and Astrophysics, MC 249-17, California Institute of Technology,
Pasadena CA 91125, USA.}
\author[0000-0002-5619-4938]{Mansi~M.~Kasliwal}
\affiliation{Cahill Center for Astronomy and Astrophysics, MC 249-17, California Institute of Technology,
Pasadena CA 91125, USA.}
\author[0000-0003-2705-4941]{Lauren Rhodes}\affiliation{Trottier Space Institute at McGill, 3550 Rue University, Montreal, Quebec H3A 2A7, Canada.}
\affiliation{Department of Physics, McGill University, 3600 Rue University, Montreal, Quebec H3A 2T8, Canada.}
\author[0000-0003-1680-7936]{Oleg M. Smirnov}\affiliation{Centre for Radio Astronomy Techniques and Technologies, Department of Physics \& Electronics, Rhodes University, Makhanda, 6140, South Africa.}
\affiliation{South African Radio Astronomy Observatory, Cape Town, 7925, South Africa}
\affiliation{Institute for Radioastronomy, National Institute of Astrophysics (INAF IRA), Via Gobetti 101, 40129 Bologna, Italy }
\author[0000-0002-9190-662X]{Davide Lazzati}
\affiliation{Oregon State University, Department of Physics, 301 Weniger Hall, Oregon State University, Corvallis, OR 97331, USA}
\author[0000-0001-8503-6958]{Joeri van Leeuwen}
\affiliation{ASTRON, the Netherlands Institute for Radio Astronomy, Oude Hoogeveensedijk 4,7991 PD Dwingeloo, The Netherlands.}
\author[0000-0001-9434-3837]{Adam Deller}
\affiliation{Centre for Astrophysics and Supercomputing, Swinburne University of Technology, Mail Number H74, PO Box 218, Hawthorn, VIC 3122, Australia.}
\author[0000-0001-8125-5619]{Pikky Atri}
\affiliation{ASTRON, the Netherlands Institute for Radio Astronomy, Oude Hoogeveensedijk 4,7991 PD Dwingeloo, The Netherlands.}
\affiliation{Department of Astrophysics/IMAPP, Radboud University, P.O. Box 9010, 6500 GL, Nijmegen, The Netherlands.}
\author[0000-0002-2483-5569]{Tanazza Khanam}
\affiliation{Department of Physics and Astronomy, Rice University, Main Street,
Houston, TX 77005, USA.}

\begin{abstract}
On 2025 August 18, the LIGO–Virgo–KAGRA collaboration reported S250818k, a sub-threshold gravitational-wave (GW) candidate consistent with a binary neutron star (NS) merger potentially involving a sub-solar–mass NS. Optical follow-up by the Zwicky Transient Facility identified AT2025ulz, a transient temporally coincident with the GW trigger that initially resembled a kilonova but was later classified as a young stripped-envelope Type IIb supernova (SN), dubbed SN\,2025ulz. A key question is whether SN\,2025ulz harbors fast, possibly collimated, non-thermal ejecta indicative of a central engine, as invoked in ``superkilonova'' scenarios linking sub-solar–mass NSs to accretion-disk fragmentation or core fission. We present early-to-late-time multi-band radio observations of SN\,2025ulz obtained with the Karl G. Jansky Very Large Array as part of the JAGWAR program, complemented by observations with the upgraded Giant Metrewave Radio Telescope and MeerKAT. We detect a faint but significant radio counterpart to SN\,2025ulz at $6-10$\,GHz. The data are consistent with non-thermal emission from SN ejecta interacting with circumstellar material, favoring a compact progenitor and relatively fast ejecta akin to those of Type cIIb SNe. Our data are also consistent with emission from an off-axis jet peaking at $\sim 50–100$ days after the GW trigger. Overall, our radio detection is compatible with a superkilonova scenario and would motivate future systematic multi-wavelength follow-up of core-collapse events coincident with sub-solar NS GW candidates, should the association between S250818k and SN\,2025ulz be supported by offline GW analyses.
\end{abstract}


\section{Introduction} \label{sec:intro}
The joint detection of gravitational waves (GWs) and an electromagnetic (EM) counterpart from the binary neutron star (NS) merger GW170817 \citep{2017PhRvL.119p1101A,2017ApJ...848L..12A} during the second observing run (O2) of Advanced LIGO and Advanced Virgo, firmly established multi-messenger astronomy as a new way to explore the cosmos \citep[e.g.,][]{Alexander_2017,Andreoni_2017,Evans_2017,2017Sci...358.1579H,2017Sci...358.1559K,2017Natur.551...75S,Troja_2017,Troja_2020,Valenti_2017,Soares_Santos_2017,Villar_2017,2018ApJ...861L..10C,2018PhRvL.120x1103L,Margutti_2018,2018ApJ...867...57R,Nynka_2018,Dobie_2018,Alexander_2018,Mooley_2018,Hajela_2019,2019NatAs...3..940H,Lamb_2019, Troja_2019, Balasubramanian_2021,Makhathini_2021,Balasubramanian_2022}. During the third observing run O3 of the LIGO-Virgo-KAGRA detectors \citep{2023PhRvX..13d1039A}, further progress was made with the identification of one binary NS coalescence with total mass significantly higher than previously known Galactic binary NS systems \citep{2020ApJ...892L...3A}, and the identification of NS-black hole merger candidates \citep{2021ApJ...915L...5A}. However, no definitive EM counterparts were discovered. 

The first part of the most recent LIGO-Virgo-KAGRA fourth observing run (O4) has doubled the census of compact binary coalescences from the first three observing runs \citep{2023PhRvX..13d1039A}. The Gravitational Wave Transient Catalog (GWTC-4) now contains a total of 218 compact binary coalescences with a probability $\gtrsim 50\%$ of being astrophysical \citep{2025ApJ...995L..18A,2025arXiv250818082T}. The catalog is dominated by binary black hole events, and the binary NS and NS-black hole local merger rates have been revised to $7.6-250$\,Gpc$^{-3}$\,yr$^{-1}$ for binary NS (compared to $10-1700$\,Gpc$^{-3}$\,yr$^{-1}$ in GWTC-3) and $9.1-84$\,Gpc$^{-3}$\,yr$^{-1}$ for NS-black hole systems \citep[compared to $7.8-140$\,Gpc$^{-3}$\,yr$^{-1}$ in GWTC-3;][]{2025arXiv250818083T}.

During the last part of O4, on 2025 August 18 at 01:20:06.030 UTC, the LIGO-Virgo-KAGRA Collaboration identified a compact binary merger candidate, S250818k, from the low-latency analysis of LIGO and Virgo data \citep{2025GCN.41437....1L}. The event was of relatively low significance, with a false-alarm rate of 2.1\,yr$^{-1}$. The 90\% credible sky-localization region spanned $\approx 786\,$deg$^2$, and the inferred luminosity distance was $259 \pm 74$\,Mpc. Follow-up observations by the Zwicky Transient Facility (ZTF) covered a substantial fraction of the localization area, leading to the identification of ZTF25abjmnps/AT2025ulz as the only candidate exhibiting a red color and a photometric redshift consistent with the GW distance \citep{2025GCN.41468....1A, 2025GCN.41476....1B, 2025GCN.41421....1B, 2025GCN.41489....1D, 2025GCN.41454....1G, 2025GCN.41433....1H, 2025GCN.41436....1K, 2025ApJ...995L..59K, 2025GCN.41474....1K,
2025GCN.41461....1L, 2025GCN.41492....1M, 2025GCN.41456....1M, 2025GCN.41439....1N, 2025GCN.41452....1O,      2025GCN.41480....1P, 2025GCN.41493....1S,  2025GCN.41451....1S,2025GCN.41414....1S}. 

Although a chance coincidence between S250818k and SN\,2025ulz cannot be ruled out, early optical follow-up within the first week revealed properties reminiscent of a GW170817-like kilonova \citep{2025ApJ...995L..59K}. While this set expectations for a potential gamma-ray burst (GRB) association \citep{2025GCN.41441....1D}, or for an off-axis GRB afterglow to emerge in the optical band after the fading of the kilonova-like emission \citep[as observed in GW170817 e.g.,][]{2019ApJ...883L...1F,2019ApJ...870L..15L}, subsequent observations confirmed the presence of a stripped-envelope Type IIb SN at redshift $z=0.0848$ \citep{2025GCN.41503....1A, 2025GCN.41518....1A, 2025GCN.41519....1A, 2025GCN.41532....1B, 2025GCN.41502....1B, 2025GCN.41544....1B, 2025ApJ...994L..45F,2025GCN.41535....1B, 2025GCN.41507....1F, 2025ApJ...995L..27G, 2025GCN.41540....1G, 2025GCN.41538....1K, 2025GCN.41534....1L, 2025GCN.41548....1P, 2025GCN.41504....1P, 2025GCN.41501....1S, 2025GCN.41837....1S, 2025GCN.41505....1T, 2025GCN.41506....1T,2026ApJ...996L..24Y}. Non-detections in the radio and X-ray bands up to $\sim$40 days after the GW trigger further disfavored a scenario in which the optical emission is dominated by a non-thermal GRB-like afterglow. Nevertheless, these constraints did not exclude the presence of a faint, off-axis relativistic jet \citep{2025GCN.41528....1B,2025GCN.41453....1H,2025arXiv251023728O,2025GCN.41464....1R,2025GCN.41542....1R}.

If the association between S250818k and SN\,2025ulz is real, a key aspect of it is that the GW signal is consistent with a merger involving sub-solar–mass NSs. Stellar evolution models generally predict a lower mass limit of $\sim 1.2$\,M$_{\odot}$ for NSs \citep{2025PhRvL.134g1403M}, while observationally NSs with masses as low as $1.02\pm0.17$\,M$_{\odot}$ have been reported \citep{2015A&A...577A.130F,2016ARA&A..54..401O}. At the same time, theoretical studies indicate that stable NSs with masses down to $\sim 0.1$\,M$_\odot$ may exist in highly neutron-rich environments, such as those produced following the collapse of a rapidly rotating star \citep{2002A&A...385..301H}. Formation channels invoking such low-mass systems e.g., through accretion-disk fragmentation or core fission in core-collapse SNe, require extreme rotation \citep{1998AstL...24..206I,2007ApJ...658.1173P,2016MNRAS.463.1642P,2002ApJ...579L..63D,2024ApJ...971L..34M,2022ApJ...941..100S,2025ApJ...991L..22C}. 

The kilonova-like early-time optical emission of SN\,2025ulz and its potential association with S250818k, resulted in a proposal to extend the so-called ``superkilonova'' model, first used to describe a collapsar producing several solar masses of heavy elements by r-process nucleosynthesis \citep{2022ApJ...941..100S}, to more broadly include any core-collapse SN that has any kilonova-like r-process nucleosynthesis inside it \citep{2025arXiv251023732K}. 

Within the superkilonova scenario, the detection of a radio afterglow from S250818k/SN\,2025ulz would provide supporting evidence for this model, linking SN\,2025ulz to a rapidly rotating progenitor capable of launching mildly- to highly-relativistic outflows (ranging from cocoons to GRB-like jets). As demonstrated in the case of GW170817, as well as in the context of long GRBs and of broad-lined Type Ic SNe, an off-axis non-thermal afterglow can be effectively probed at radio wavelengths \citep[e.g.,][]{1998Natur.395..663K,Patat_2001,2003ApJ...599..408B,2004Natur.430..648S,2010Natur.463..513S,2006ApJ...639..331G,2013ApJ...778...63H,2013ApJ...778...18M,2016ApJ...830...42C,Alexander_2017,2017ApJ...847...54C,2018ApJ...863...32D,Dobie_2018, Mooley_2018,2020ApJ...902...86H,2020ApJ...893..132H,Makhathini_2021}. Indeed, one may expect the optical afterglow to remain  undetected while the emission is dominated by kilonova or SN light at optical wavelengths \citep[][]{2019ApJ...883L...1F,2019ApJ...870L..15L}. On the other hand, radio emission can also arise from the interaction of SN ejecta with circumstellar material (CSM). In fact, CSM-interaction is a radio-emission mechanism commonly observed in Type IIb SNe \citep[e.g.,][]{1982ApJ...259..302C,Chevalier_1998,1998ApJ...509..861F,2003ApJ...592..900S,2004MNRAS.349.1093R,2007ApJ...671.1959W, Chevalier_2010,2018SSRv..214...27C,2016ApJ...818..111K,2021ApJ...923L..24S,2024MNRAS.534.3853R,Chandra2025,2025PASA...42...50S}, and can be observed in stripped-envelope SNe as well \citep[e.g.,][]{2003ApJ...599..408B, 2006ApJ...638..930S,2014ApJ...782...42C,  2016ApJ...830...42C, 2023ApJ...953..179C,2025arXiv251208822O}. Radio observations can be used to test whether the observed emission is more likely related to slow ejecta interacting with a dense CSM, or rather with weaker CSM interaction of a faster ejecta. The last would be more in line with what may be expected in the context of an extended superkilonova model, given the need for extreme rotation.

Overall, recognizing that the fate of fast-rotating collapses can be rich and that the properties of their ejecta is likely to be diverse \citep[e.g.,][]{2001ApJ...550..410M,2006ARA&A..44..507W,2008ApJ...676.1130M,2012ApJ...749...91F,2017MNRAS.472..616S,2022MNRAS.516.2252O,2024ApJ...971L..34M}, any evidence for non-thermal radio emission related to the presence of fast ejecta in SN\,2025ulz (though not necessarily in the form of a GRB-like jet) would support the case for more investigation into the extended superkilonova hypothesis, especially by ensuring that extensive multi-wavelength follow up of core-collapses found in association with sub-solar NS GW candidates is carried out in the future observing runs of the LIGO-Virgo-KAGRA detectors. 

Motivated by the above considerations, here we present deep radio observations of the SN\,2025ulz field carried out with the Karl G. Jansky Very Large Array (VLA) through our ``Jansky VLA mapping of Gravitational Waves as Afterglows in Radio'' (JAGWAR) program. As we describe in detail in Section \ref{sec:obs}, we find evidence for a faint radio counterpart associated with SN\,2025ulz, peaking around $50-100$ days since the S250818k trigger at $6-10$\,GHz. In Section \ref{sec:modeling}, we describe the constraints that our radio observations place on both the SN-CSM interaction model and an off-axis GRB afterglow scenario. Finally, in Section \ref{sec:conclusion} we summarize and conclude.

\begin{figure}
\centering
\vbox{
\includegraphics[width=0.45\textwidth]{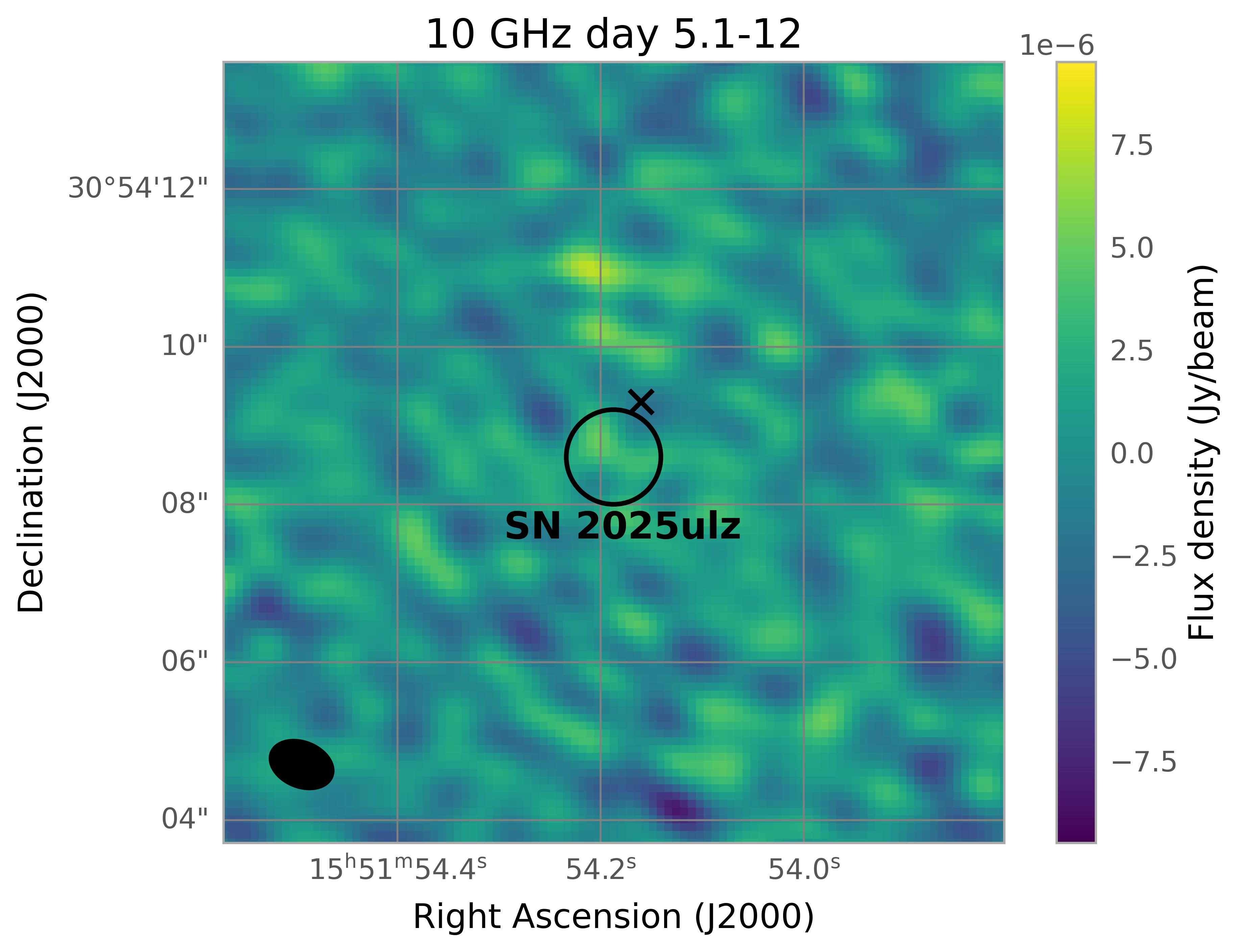}
\includegraphics[width=0.45\textwidth]{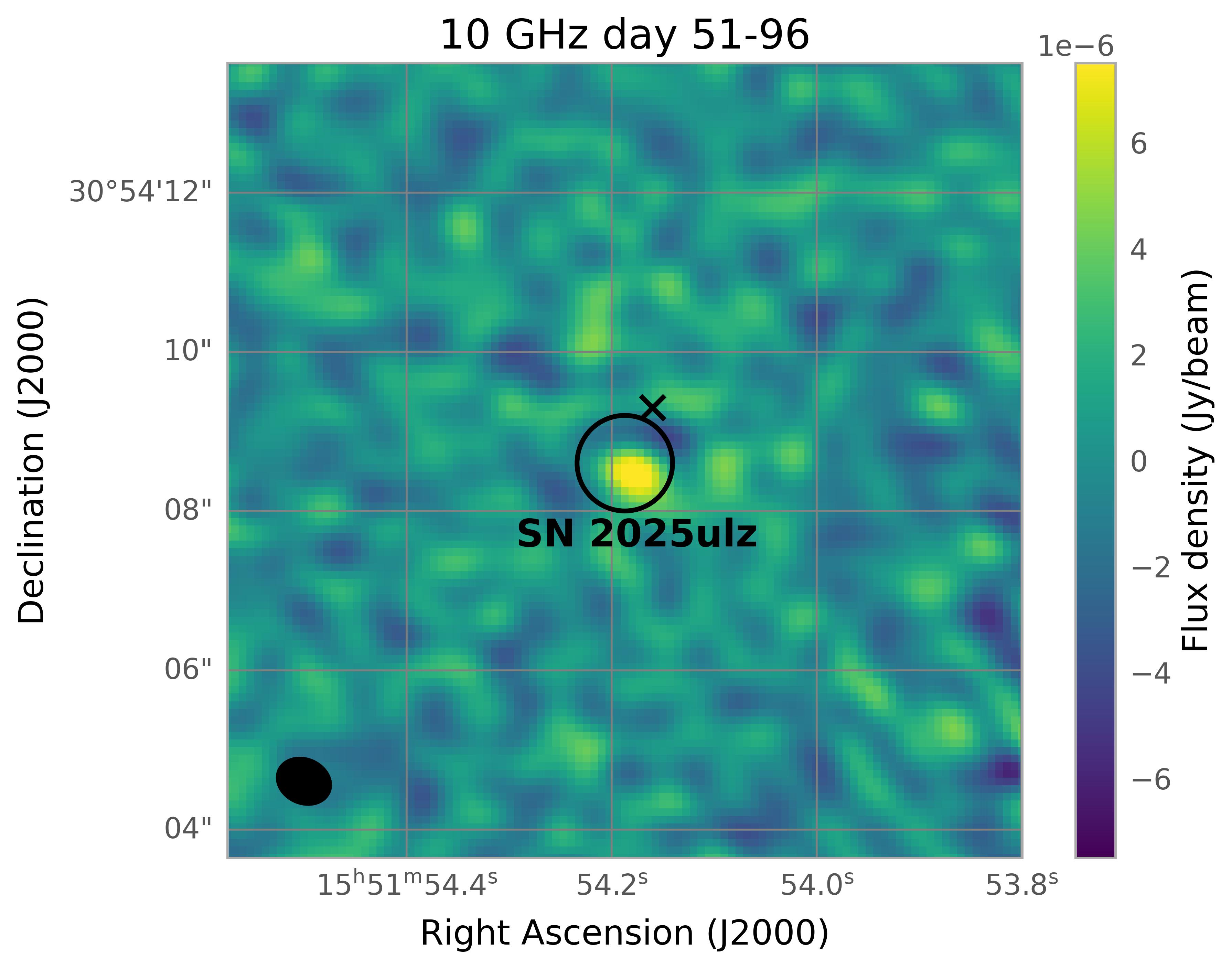}
\includegraphics[width=0.46\textwidth]{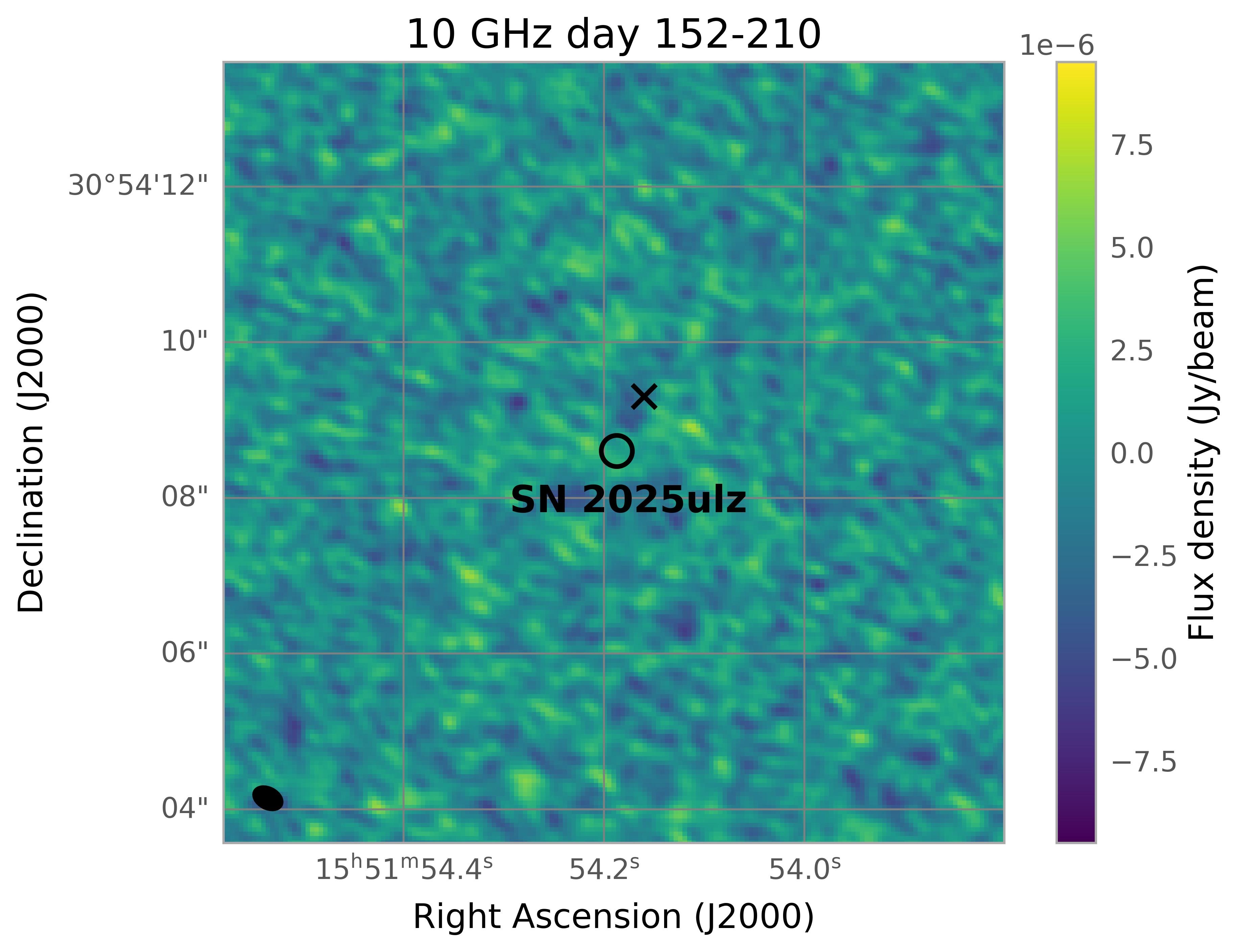}}
\caption{VLA images of the SN\,2025ulz field at $\approx 10$\,GHz. In all panels, we show a black circle centered on the HST position of SN\,2025ulz with radius equal to the nominal FWHM of the VLA synthesized beam (0.6\,\arcsec in the top and middle panels; 0.2\,\arcsec in the bottom panel). The actual VLA beams are plotted with black ellipses in the bottom-left corner of each panel. The optical position of the host galaxy SDSS
J155154.16+305409.3 is marked with a black cross. Maps' colors span the $\pm5\times$\,RMS flux density range. The radio counterpart of SN\,2025ulz detected at the $\approx 6\sigma$ level in the central panel. The counterpart is not detected in preceding observations (top panel) and fades below detection level in the later observations (bottom panel).}
\label{fig:Xband_discovery}
\end{figure}

\section{Observations and data reduction} \label{sec:obs}
We carried out multi-band radio observations of the SN\,2025ulz field with the VLA as part of our JAGWAR program. We also observed the same field with the upgraded Giant Metrewave Radio Telescope (uGMRT) through a complementary program. For completeness, we additionally re-analyzed MeerKAT observations of the field obtained as part of an independent program \citep{2025GCN.41500....1B,2025GCN.41594....1B,2025GCN.42032....1B}. The results of our observations and data reduction are reported in Table \ref{tab:a}, which also includes radio observations from other works \citep{oconnor2025at2025ulzs250818kdeepxray}. In what follows, we provide details on how we reduced each data set. 

\subsection{uGMRT}

The uGMRT observed the field of SN\,2025ulz  starting on 2025 August 26.57 UT for 2 hrs total \citep[this includes calibration overhead;][]{2025GCN.41577....1B}.  The observations were carried out in total intensity mode (Stokes I) in band 5 ($1000-1450$\,MHz)   with an integration time of 10 s. The calibrator sources 3C286 and J1609$+$266 were used for flux plus bandpass calibration, and phase calibration, respectively. 

The data were analyzed using Common Astronomy Software Applications package \citep[\texttt{CASA,}][]{2007ASPC..376..127M}. Initially the 
calibration and imaging pipeline for the uGMRT interferometric data, CAsa Pipeline-cum-Toolkit for Upgraded GMRT data REduction \citep[\texttt{CAPTURE,}][]{2021ExA....51...95K}, was run. 
The flagged and calibrated data were closely inspected, and further flagging and calibration were carried out manually until the data quality looked satisfactory.  A few rounds of phase self-calibration were performed. Our final image had a resolution of $2.5\arcsec\times2.0\arcsec$. We did not detect any radio emission at the position of SN\,2025ulz.  Our 3$\sigma$ upper limit is reported in Table \ref{tab:a}.

\subsection{MeerKAT}

MeerKAT is a radio interferometer in the Karoo Desert, South Africa made up of  64, 13.5-meter dishes. We report on two observations obtained through an approved open call proposal SCI-20241101-GB-01 \citep[PI: Bruni;][]{2025GCN.42032....1B} and made at a central frequency of 3062\,MHz with a bandwidth of 875\,MHz. The observations started at 16:32 on 2025 August UT, and at 14:22 on 2025 August 25 UT. PKS B1934$-$638 (J1939$-$6342) was used as flux density and bandpas calibrator, and  J1609+2641 as phase calibrator.

Both epochs were reduced with the \texttt{oxkat} package, specifically designed for semi-automatic processing of MeerKAT data \citep{2020ascl.soft09003H}. Persistent RFI was removed from the calibrator fields before performing bandpass calibration and flux density scaling using J1939$-$6342. All first-generation calibration steps were performed in \texttt{CASA} \citep{2007ASPC..376..127M}.  Imaging was performed with \textsc{wsclean} \citep[Version 2.5,][]{2014MNRAS.444..606O}, using a Briggs weighting with a robust parameter of -0.7. We also performed a single round of phase-only self-calibration after which the target field was re-imaged.

In both epochs, we detect a resolved source at the coordinates of SN\,2025ulz which we associate with emission dominated by host galaxy light \citep[see also][]{2025GCN.41666....1R}.  We measure the component to be ($9\arcsec\pm1\arcsec)\times(2.9\pm0.2)\arcsec$ with a position angle of $(16.1\pm0.1)\arcdeg$. For comparison, the clean beam size is a 3.5\arcsec$\times$3.5\arcsec. Our peak flux density measurements are reported in Table \ref{tab:a}.

\begin{figure*}
   \centering
\includegraphics[width=\textwidth]{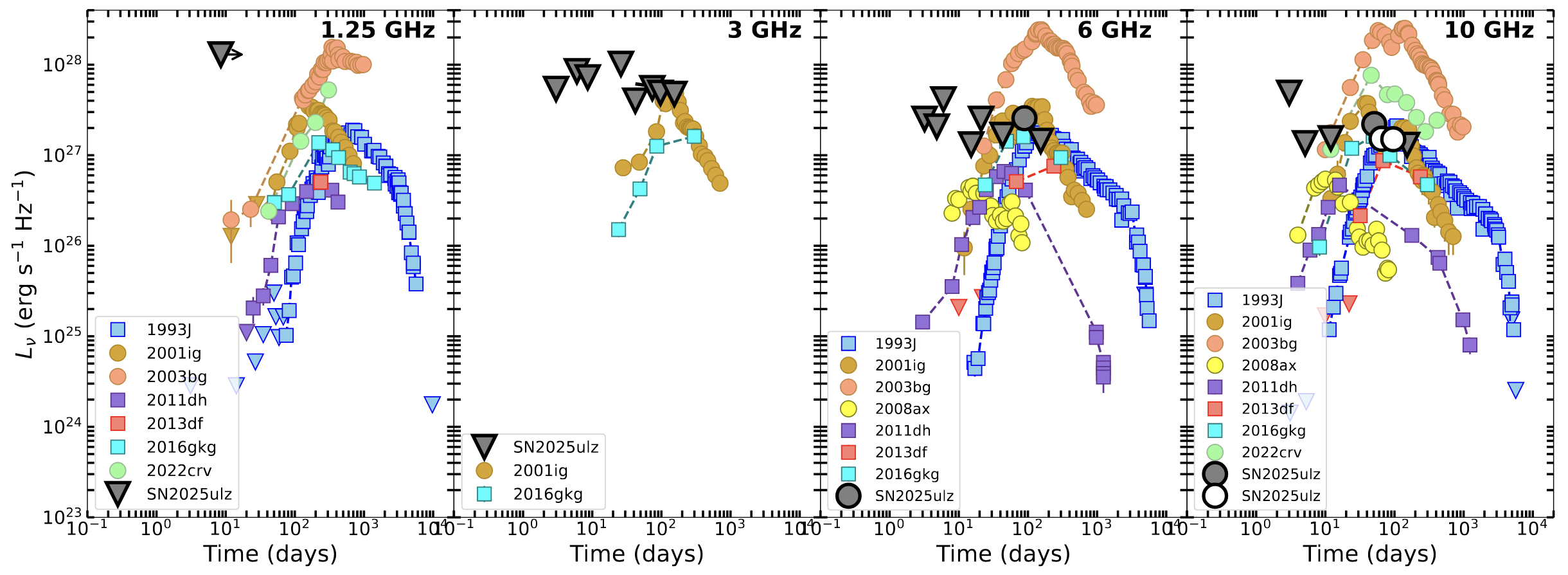}
    \caption{The radio light curves of SN 2025ulz compared to that of Type IIb SNe. For SN 2025ulz, we plot $5\sigma$ detections with dots, detections at $\ge 3\sigma$ (but $<5\sigma$) with open circles, and upper limits (including detections dominated by host galaxy light) with downward triangles (Table \ref{tab:a}).  The data for SNe Type IIb are taken from \citet{Chandra2025} and references therein. See Section \ref{sec:supernova} for discussion. }
    \label{fig:sn}
\end{figure*}

\subsection{Jansky VLA}
We observed the field of SN\,2025ulz with the VLA in CnB, B, and A configurations over $3–15$\,GHz as part of the JAGWAR program (PI: Corsi). Observations were conducted in $\sim$3\,hr blocks per frequency (including calibration). The first epoch began on 2025 August 21 UT ($\sim$3 days post S250818k), and the last on 2026 March 17 ($\sim$211 days post-trigger). We used J1602+3326 as phase calibrator and 3C286 for flux density and bandpass calibration. All observations are listed in Table \ref{tab:a}.

Data calibration and imaging were performed using the VLA calibration and imaging pipeline \citep{2020ASPC..527..571K} in \texttt{CASA} \citep{2022PASP..134k4501C}. With this pipeline, after calibration and automated flagging, imaging is performed using the task \texttt{tclean} with auto-masking, cleaning down to the $4\sigma$ level using. Briggs weighting with \texttt{robust}=0.5 is adopted as a default for continuum imaging. \texttt{Nterms}=2 is used for large bandwidth observations, and self-calibration solutions are applied to the data when the self-calibration is successful.
After automated calibration and imaging, all data were also manually inspected for the presence of potential RFI effects. Particularly in S-band, flagging and re-imaging was performed as necessary to test the robustness of our results. In all cases, imaging manually using \texttt{tclean} after inspecting the data yielded results consistent with the automated calibration and imaging pipeline, within errors. 

Fluxes reported in Table \ref{tab:a} are measured using the  \texttt{imstat} tool in \texttt{CASA}. For each VLA epoch, we report the maximum flux density found within a circular region centered on the Hubble Space Telescope position on SN\,2025ulz ($\alpha$=15$^{\rm h}$51$^{\rm m}$54.187$^{\rm s}$, $\delta=$30\arcdeg54\arcmin08.602\arcsec), with radius equal to the nominal VLA synthesized beam at the configuration and frequency setup of each observation. The root-mean-square (RMS) of the noise is estimated from each image using a circular region of radius $\approx 10\times$ the nominal FWHM of the synthesized beam centered on the position of SN\,2025ulz. We also verified that the noise RMS estimated in this region is consistent with the RMS estimated using a circular region of the same size free of sources. 

A faint radio counterpart to SN\,2025ulz is detected at $>3\sigma$ in X-band ($\approx 10$\,GHz; Figure \ref{fig:Xband}) and C-band ($\approx 6$\,GHz; Figure \ref{fig:Cband}). To improve the robustness of the detection, we concatenate X-band observations obtained pre-, during, and post-discovery (Figure \ref{fig:Xband_discovery}). At this frequency, host galaxy contamination, prominent in S-band ($\approx 3$\,GHz; Figure \ref{fig:Sband}), is minimized. Imaging of the combined data yields a $\approx 6\sigma$ detection between days 51--96 post-trigger (Figure \ref{fig:Xband_discovery}, central panel). The inferred radio position is $\alpha=15^{\rm h}51^{\rm m}54.180^{\rm s}$, $\delta=30\arcdeg54\arcmin08.45\arcsec$, with an uncertainty of $\approx 0.11\,\arcsec$ (dominated by a $0.10\,\arcsec$ systematic added in quadrature). This is offset by $\approx 0.18\,\arcsec$ from the HST position, which itself has an uncertainty of $\approx 0.10\,\arcsec$.

We finally note a possible hint of brightening in S-band at day 74 (Figure \ref{fig:Sband}), potentially due to emission from the SN\,2025ulz radio counterpart; however, contamination from the host galaxy prevents a statistically significant disentanglement of this contribution.

\begin{figure}
    \centering \includegraphics[width=0.45\textwidth,height=0.5\textwidth]{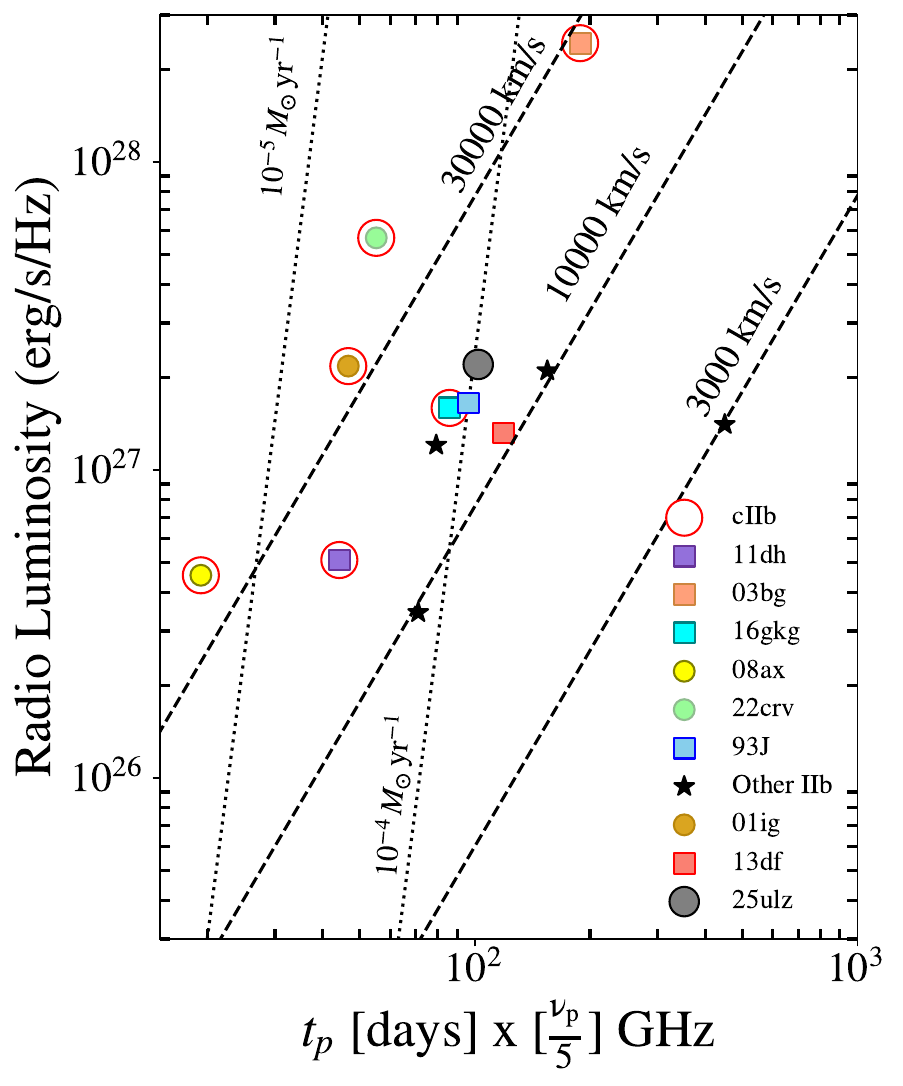}
    \caption{A Chevalier diagram (peak radio luminosity vs peak time) with the 10\,GHz detection (likely peak) of SN\,2025ulz (gray dot) compared to that of other Type IIb SNe (as indicated in the legend). Lines of constant shock velocity and mass-loss rate are shown. Type cIIb explosions thought to be linked to compact progenitors are highlighted with open red circles. See Section \ref{sec:supernova} for discussion.}
    \label{fig:Chevalier}
\end{figure}

\begin{figure}
    \centering
    \includegraphics[width=0.49\textwidth]{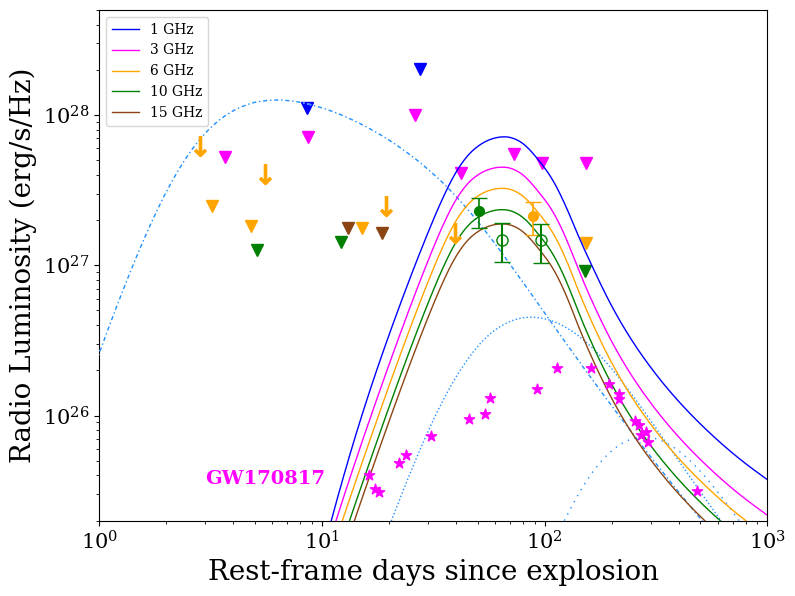}
    \caption{Radio observations of SN\,2025ulz. We show upper limits (downward triangles), $\ge 5\sigma$ detections (dots), and $3\sigma$–$4.9\sigma$ detections (open circles). Downward triangles also mark $\ge 3\sigma$ emission in the search region likely due to host galaxy light (see Table \ref{tab:a}). Additional $6$\,GHz upper limits from \citet{oconnor2025at2025ulzs250818kdeepxray} are included (arrows). For comparison, we plot the $3$\,GHz light curve of GW170817 \citep{Makhathini_2021} (magenta stars). Light blue dot-dashed, dotted, and densely dotted curves show models from \citet{oconnor2025at2025ulzs250818kdeepxray}, while solid curves are our models computed with \texttt{PyBlastAfterglowMag} \citep{2024arXiv240916852N, 2023MNRAS.tmp..259N}. See Section \ref{sec:grb} for discussion.
    }
    \label{fig:Afterglow models}
\end{figure}
\section{Radio modeling} 
\label{sec:modeling}

Radio emission from explosive transients associated with core collapse SNe and/or GRBs probes the fastest outflow components, as well as the density and structure of the CSM \citep[e.g.,][and references therein]{1982ApJ...259..302C,Chevalier_1998,1998ApJ...509..861F,1998Natur.395..663K,1998ApJ...497L..17S,1999ApJ...516..788W,2000ApJ...536..195C,2003ApJ...599..408B,2003ApJ...586..356Z,Chevalier_2010,2015MNRAS.450.1430H,2017Sci...358.1579H,2017ApJ...834...28N,2018SSRv..214...27C,2018ApJ...856...56C,2018ApJ...863...32D,2018PhRvL.120x1103L}. In what follows, we describe the constraints that our observations set on the radio emitting ejecta of SN\,2025ulz in the context of two main models:  (i) SN ejecta interacting with the CSM (Section \ref{sec:supernova}); (ii) A rapidly-rotating progenitor powering a SN plus a fast, collimated ejecta (Section \ref{sec:grb}).

\subsection{CSM interaction in SN\,2025ulz}
\label{sec:supernova}
Among stripped-envelope core-collapse explosions, Type IIb SNe retain only trace amounts of hydrogen and provide an important bridge between H-rich SNe and more highly stripped events, including the rare broad-lined Ic SNe associated with long GRBs. The interaction between the expanding SN ejecta and the CSM can power broadband  emission. As fast ejecta ($\sim 10{,}000$–$50{,}000$\,km\,s$^{-1}$) collide with slower CSM ($\sim 10$–$10,00$\,km\,s$^{-1}$), a forward shock propagates into the CSM while a reverse shock travels back into the ejecta. These shocks convert kinetic energy into thermal and non-thermal radiation, with relativistic electrons accelerated in magnetized shocks producing synchrotron radio emission \citep{1979rpa..book.....R}.

In Figure \ref{fig:sn} we compare the radio light curves of Type IIb SNe from \citet{Chandra2025} with SN\,2025ulz. At 6 and 10\,GHz, SN\,2025ulz evolves more rapidly than most Type IIb SNe, with a fast rise and decline indicative of a compact progenitor, similar to SN\,2011dh. Type IIb SNe are broadly divided into compact (cIIb) and extended (eIIb) classes \citep{Chevalier_2010}: cIIb events resemble SNe Ib/c, showing faster evolution, weaker CSM interaction, and higher ejecta velocities, and are linked to Wolf–Rayet stars or stripped binaries, whereas eIIb SNe arise from extended supergiants. Our radio data favor a compact progenitor for SN\,2025ulz, consistent with optical constraints \citep{2025ApJ...995L..59K}.

In stripped-envelope core-collapse SNe (Types IIb/Ib/Ic), synchrotron self-absorption (SSA) is typically the dominant radio absorption mechanism. Under this assumption, Figure \ref{fig:Chevalier} shows a Chevalier diagram (peak luminosity vs. peak time; \citealt{Chevalier_1998}) including the 10\,GHz detection of SN\,2025ulz (likely near peak) compared to other Type IIb SNe. The diagram is constructed following \citet{2012ApJ...752...78S}, assuming a wind velocity of 500\,km\,s$^{-1}$ and including lines of constant shock velocity. Interpreted in this framework, the radio emission from SN\,2025ulz implies ejecta expanding at $\sim 15{,}000$\,km\,s$^{-1}$ into a CSM shaped by a progenitor with mass-loss rate $\dot{M}\sim 10^{-4}$\,M$_{\odot}$\,yr$^{-1}$ (for $\epsilon_B=\epsilon_e=0.1$ and $p=3$).

\subsection{An off-axis jet afterglow accompanying SN\,2025ulz}
\label{sec:grb}
The detection of a radio afterglow from S250818k/SN\,2025ulz would provide support to a scenario involving  a rapidly rotating progenitor capable of launching relativistic outflows. 

In Figure \ref{fig:Afterglow models}, we compare the radio data of SN\,2025ulz with 
model light curves of GRB afterglows potentially compatible with the observations. The dot-dashed, dotted, and densely dotted light blue curves in Figure \ref{fig:Afterglow models}  are models that were considered viable and not yet excluded in the analysis presented in \citet{oconnor2025at2025ulzs250818kdeepxray}. To model our VLA observations, we use \texttt{PyBlastAfterglowMag} \citep{2024arXiv240916852N, 2023MNRAS.tmp..259N} and set the power-law index of the electron energy distribution to $p =2.01$, and the fraction of energies going into accelerated electrons and magnetic fields to $\epsilon_e =0.01$ and $\epsilon_B=0.001$, respectively. Solid curves in Figure \ref{fig:Afterglow models} are model light curves for a Gaussian jet with isotopic equivalent energy along the jet axis of $E=1.8\times10^{53}$erg and a  jet core angle of $\theta_{core}=4.01\,\arcdeg$. We note that this jet energy is most compatible with that of long GRBs associated with collapses of highly-rotating massive stars \citep{2017ApJ...837..119A,2019MNRAS.488.5823L}, and higher than the kinetic energies of short GRBs \citep{2014ARA&A..52...43B}. The jet is observed at an angle $\theta_{obs} = 32.1\,\arcdeg$, and expands in a constant density interstellar medium (ISM) with particle number density of $n_{\rm ISM} =1.4$\,cm$^{-3}$. 

While the off-axis GRB jet model provides a plausible interpretation of the radio counterpart to SN\,2025ulz, the inferred parameters should be regarded as indicative given degeneracies that cannot be resolved with the current dataset. For example, for $p\approx 2$–3 and neglecting SSA, extrapolation from the 10\,GHz detection (Figure \ref{fig:Xband}) predicts $\sim 12$–15\,$\mu$Jy at 6\,GHz and $\sim 17$–30\,$\mu$Jy at 3\,GHz at $\sim$70 days, consistent with our measurements within uncertainties and host contamination (Table \ref{tab:a}). In the model shown in Figure \ref{fig:Afterglow models}, we assume the 3\,GHz emission is host-dominated and adopt $p=2.01$, which underpredicts the observed flux at this frequency. Larger values of $p$ would instead imply a greater contribution from SN\,2025ulz at 3\,GHz.

\section{Summary and Conclusion}
\label{sec:conclusion}
We presented early-to-late-time multi-band radio observations of SN\,2025ulz, a Type IIb SN that has emerged as a potential EM counterpart to the low-significance GW triggger S250818k.
Our VLA observations reveal a faint, but significant radio counterpart to SN\,2025ulz at $6-10$\,GHz. In the CSM-interaction scenario, which is motivated by the classification of SN\,2025ulz as a Type IIb SN, our radio data (light curve evolution and implied radio ejecta speed) point to properties most similar to that of cIIb events associated with compact progenitors, lower level of CSM interaction, and faster ejecta speeds compared to Type eIIb SNe. In the jet afterglow scenario, which is motivated by the similarity between the SN\,2025ulz early optical emission and that of the GW170817 kilonova, our radio observations are consistent with an off-axis relativistic jet whose emission peaks on $\sim$50–100 day timescales. 

Our radio observations of SN\,2025ulz provide key insight into the nature of this transient. If the SN\,2025ulz/S250818k association is genuine, the radio detection is broadly consistent with the proposed superkilonova scenario. However, the faintness of the counterpart and host-galaxy contamination prevent a unique interpretation, with both CSM interaction from fast ejecta and an off-axis jet remaining viable.
As in GW170817, deep radio follow-up of GW alerts has proven essential to probe ejecta components inaccessible at optical wavelengths, with the sensitivity, angular resolution, and frequency coverage of the VLA playing a critical role. Despite uncertainties in the association, SN\,2025ulz is among the most compelling GW–EM candidates of O4. Systematic radio follow-up of future GW events, particularly those involving candidate sub-solar-mass NS mergers, will be key to testing whether rapid rotation can produce sub-solar NSs and is linked to relativistic jet launching or systematically faster SN ejecta.

\begin{acknowledgements}
\small T.O. and A.C. acknowledge support from the National Science Foundation (NSF) via grant AST-2431072. K.P.M. and D.Y. acknowledge support from Anusandhan National Research Foundation grant ANRF/ECRG/2024/005949/PMS. R.B.W. is supported by the National Science Foundation Graduate Research Fellowship Program under Grant number 2234693 and acknowledges support from the Virginia Space Grant Consortium. L.R. acknowledges funding from the Natural Sciences and Engineering Research Council of Canada (NSERC) Arthur B. McDonald Fellowship and Discovery Grant programs, the Canada Research Chairs (CRC) program, the Fondes de Recherche Nature et Technologies (FRQNT), the Centre de recherche en astrophysique du Québec (un regroupement stratégique du FRQNT), the Trottier Space Institute at McGill and the AstroFlash research group. The AstroFlash research group at McGill University, University of Amsterdam, ASTRON, and JIVE is supported by: a Canada Excellence Research Chair in Transient Astrophysics (CERC-2022-00009); an Advanced Grant from the European Research Council (ERC) under the European Union’s Horizon 2020 research and innovation programme (`EuroFlash’; Grant agreement No. 101098079); an NWO-Vici grant (`AstroFlash’; VI.C.192.045); an NSERC Discovery Grant (RGPIN-2025-06681); an ERC Starting Grant (`EnviroFlash’; Grant agreement No. 101223057); and an NWO-Veni grant (VI.Veni.222.295). P.A. is supported by the WISE fellowship program, which is financed by The Dutch Research Council (NWO). The National Radio Astronomy Observatory and Green Bank Observatory are facilities of the U.S. National Science Foundation operated under cooperative agreement by Associated Universities, Inc. We thank the staff of the GMRT that made these observations possible. GMRT is run by the National Centre for Radio Astrophysics of the Tata Institute of Fundamental Research.. We thank Xander Hall for providing the reference HST position of SN\,2025ulz. We thank Avery Eddins and Steven Myers for providing feedback on this work.
\end{acknowledgements}

\begin{table*}
    \begin{center}
    \caption{Radio observations of the SN\,2025ulz field. We report from left to right: telescope name; array configuration; observing epoch in days since the GW trigger time (MJD 60905.0556); central observing frequency; maximum flux density (if $\ge 3\times$RMS, upper-limit otherwise) measured in a circular region centered on the position of SN\,25025ulz with radius equal to the nominal FWHM of the synthesized beam; noise RMS in a region with radius $10\times$ the FWHM of the nominal synthesized beam; nominal FWHM of the synthesized beam (for mixed VLA configurations, we report the beam size of the most extended configuration); radio peak angular offset and position angle (PA) from the HST position of SN\,2025ulz; a flag for the likely origin of the measured radio flux (H for host galaxy dominated, AT for SN\,2025ulz, and N for noise fluctuation); Project code; Project PI. \label{tab:a}}
    \begin{tabular}{ccccccccccc}
    \hline 
    \hline
    Tel & Config.  & Epoch & $\nu$ & Max$F_{\nu}$ & RMS & Syn.Beam & Offset,PA & Flag & ProjectCode & PI \\
     & &  (day) & (GHz) & ($\mu$Jy) & ($\mu$Jy) & ($\arcsec$) & ($\arcsec$,$\arcdeg$) & & &  \\

    \hline
    MeerKAT & $\ldots$  & 27.6 & 0.82 & $208\pm21$ & 18 & 13.9 & 0.16, 91 & H &SCI-20241101-GB-01 & Bruni\\
    \hline 
        uGMRT & $\ldots$  & 8.5 & 1.3 & $< 63$ & 21 & 2.5 &  & --& 48\_096 & Chandra\\
        MeerKAT & $\ldots$  & 27.5 & 1.3 & $114\pm12$ & 11 & 6.5 & 1.4, 220 & H &SCI-20241101-GB-01 & Bruni\\
        \hline 
        MeerKAT & $\ldots$ & 3.6 & 3.0 & $30.0\pm6.6$ & 6.4 & 3.4 & 0.83, 53 & H & SCI-20241101-GB-01 & Bruni\\ 
        VLA & CnB  & 6.0 & 3.0 & $< 47 $  & -- & --  & -- & -- & 22B-275 & Troja\\ 
        MeerKAT & $\ldots$  & 8.6 & 3.0 & $40.3\pm4.8$ & 4.4 &3.4 & 1.1, 341 & H & SCI-20241101-GB-01 & Bruni\\
        MeerKAT & $\ldots$  & 26 & 3.0 & $56.7\pm6.8$ & 6.2 & 3.4 & 1.5, 6.3 &  H & SCI-20241101-GB-01 & Bruni\\
        VLA & B  & 42 & 3.2 & $22.3\pm5.0$  & 4.6 & 2.1 & 1.0, 329 & H & 22B-235 & Corsi\\
        VLA & B  & 74  & 3.2 & $30.7\pm5.4$ & 5.2 & 2.1 & 0.22, 271 &H/AT? &  22B-235 & Corsi\\
        VLA & B  & 97 & 3.2 & $27.0\pm4.7$& 4.5 & 2.1 & 1.5, 351 & H & 22B-235 & Corsi \\
        VLA & B  & 154 & 3.2 & $27.4\pm4.8$& 4.6 & 2.1 & 0.81, 319 & H & 22B-235 & Corsi \\
    \hline

        VLA & CnB  & 3.2 & 6.2 & $7.1\pm3.1$ & 3.1 & 1.0 & 0.91, 20.5 & H & 22B-235 & Corsi\\

        VLA & CnB  & 4.8 & 6.2 & $8.2\pm2.6$ & 2.6 & 1.0 & 0.45, 309 &  H & 22B-235 & Corsi\\

        VLA & CnB  &  6.0 & 6.0 & $<23$ & -- & -- & -- & -- & 22B-275 & Troja \\ 
 
        VLA & B & 15.1 & 6.1 & $8.9\pm2.6$ & 2.6 & 1.0 & 0.97, 340 & H & 22B-235 & Corsi\\
        
        VLA & B &  21 & 6.0 & $< 14$ & -- & -- & -- & -- & SC260095 & Troja \\ 

        VLA & B &  43 & 6.0 & $< 9.3$ & -- & -- & -- & -- & SC260095 & Troja \\  

        VLA & B  & 89 & 6.2 & $14.2\pm2.7$ & 2.6 & 1.0 & 0.10, 99.6 & AT & 22B-235 & Corsi\\

        VLA & B  & 153 & 6.2 & $8.0\pm2.5$ & 2.5 & 1.0 & 0.66, 22.6 & H/N & 22B-235 & Corsi\\
        VLA & A &  211  & 6.1 & $<7.8$ &  2.6 & 0.33 & --& -- & 22B-235 & Corsi\\

        \hline
        VLA & CnB  &  3.0 & 10 & $< 27$ & -- & -- & -- & -- & 22B-275 & Troja \\   
        
        VLA & CnB  & 5.1 & 9.9 & $<7.2$ & 2.5 & 0.60 & -- & -- & 22B-235 & Corsi\\

        VLA & B & 12 & 9.9 & $8.5\pm2.7$ & 2.7 & 0.60 & 0.21, 65.4 & N & 22B-235 & Corsi\\

        VLA & B  & 51 & 9.8 & $12.2\pm2.7$ & 2.6 & 0.60 & 0.13, 225 & AT & 22B-235 & Corsi\\
        VLA & B  & 65 & 9.9 & $8.4\pm2.6$ & 2.6  & 0.60 & 0.21, 247 &  AT & 22B-235 & Corsi\\
        VLA & B & 96 & 9.9 & $8.3\pm2.5$ & 2.5 & 0.60 & 0.34, 215 & AT & 22B-235 & Corsi\\
        VLA & B & 152 & 9.9 & $<7.2$ & 2.4 & 0.60 & -- & -- & 22B-235 & Corsi\\
        VLA & A   &  210 & 9.9 &  $<7.8$&   2.6 & 0.20 & -- & -- & 22B-235 & Corsi\\
        VLA & CnB--B  & 5.1--12 & 9.9 & $<5.7$ &  1.9 & 0.60 & --& -- & 22B-235& Corsi\\
VLA & B  & 51--96 & 9.9 & 9.2 & 1.5 &  0.60& 0.18, 211 &AT & 22B-235& Corsi\\
 VLA & B--A & 152--210 & 9.9 & $<5.7$ & 1.9 & 0.20 & --& --&  22B-235&Corsi \\

        \hline
        VLA & B & 13 & 15 & $<6.9$ & 2.3 & 0.42 & -- & --& 22B-235 & Corsi\\
        VLA & B & 19 & 15 & $< 7.2$ & 2.6 & 0.42 & -- & --&  22B-235 & Corsi\\

    \hline
    \end{tabular}
    \end{center}
\end{table*}

\bibliographystyle{aasjournal}
\bibliography{main}

\appendix
\begin{figure*}[!ht]
\centering
\hbox{
\includegraphics[width=0.33\textwidth]{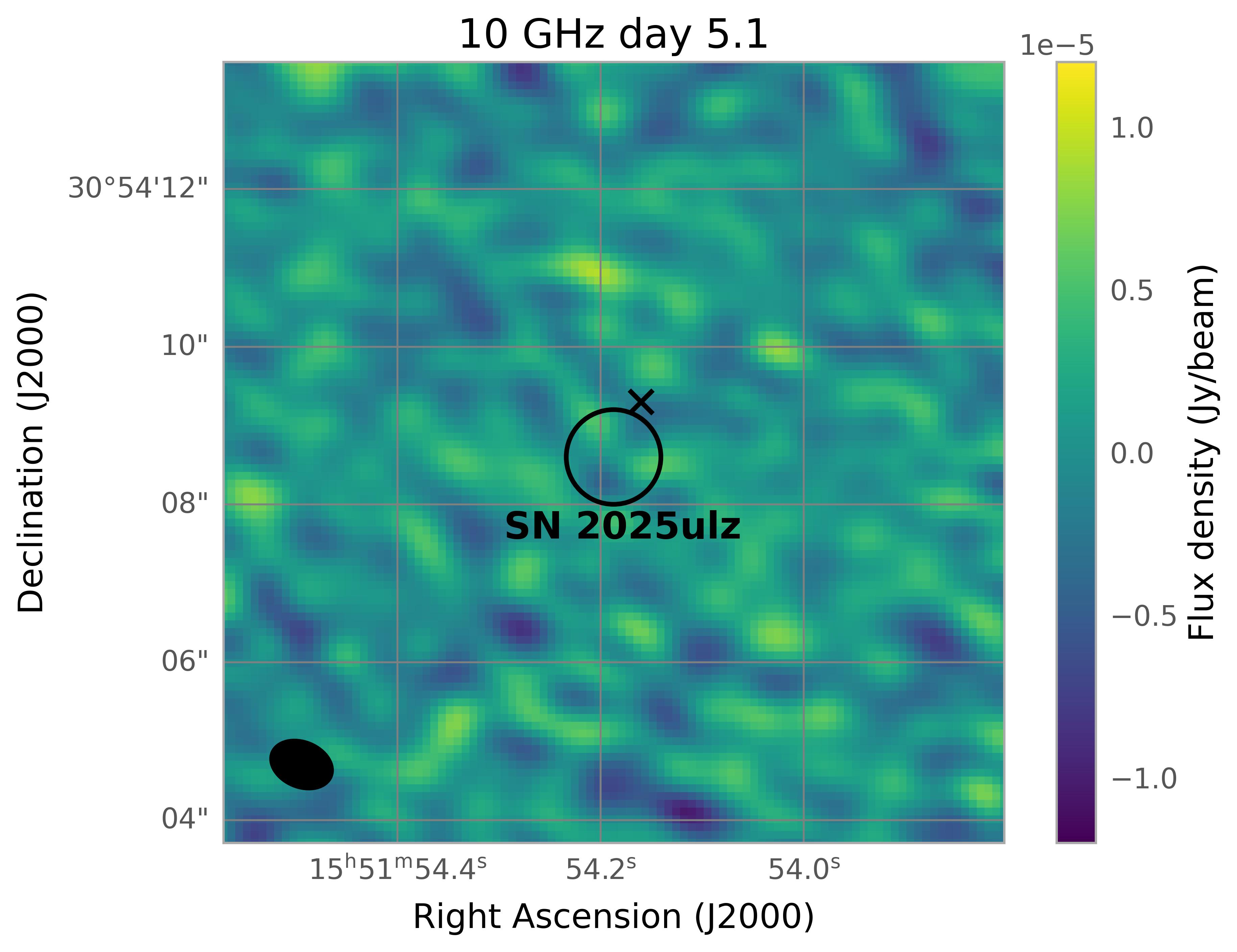}%
\includegraphics[width=0.33\textwidth]{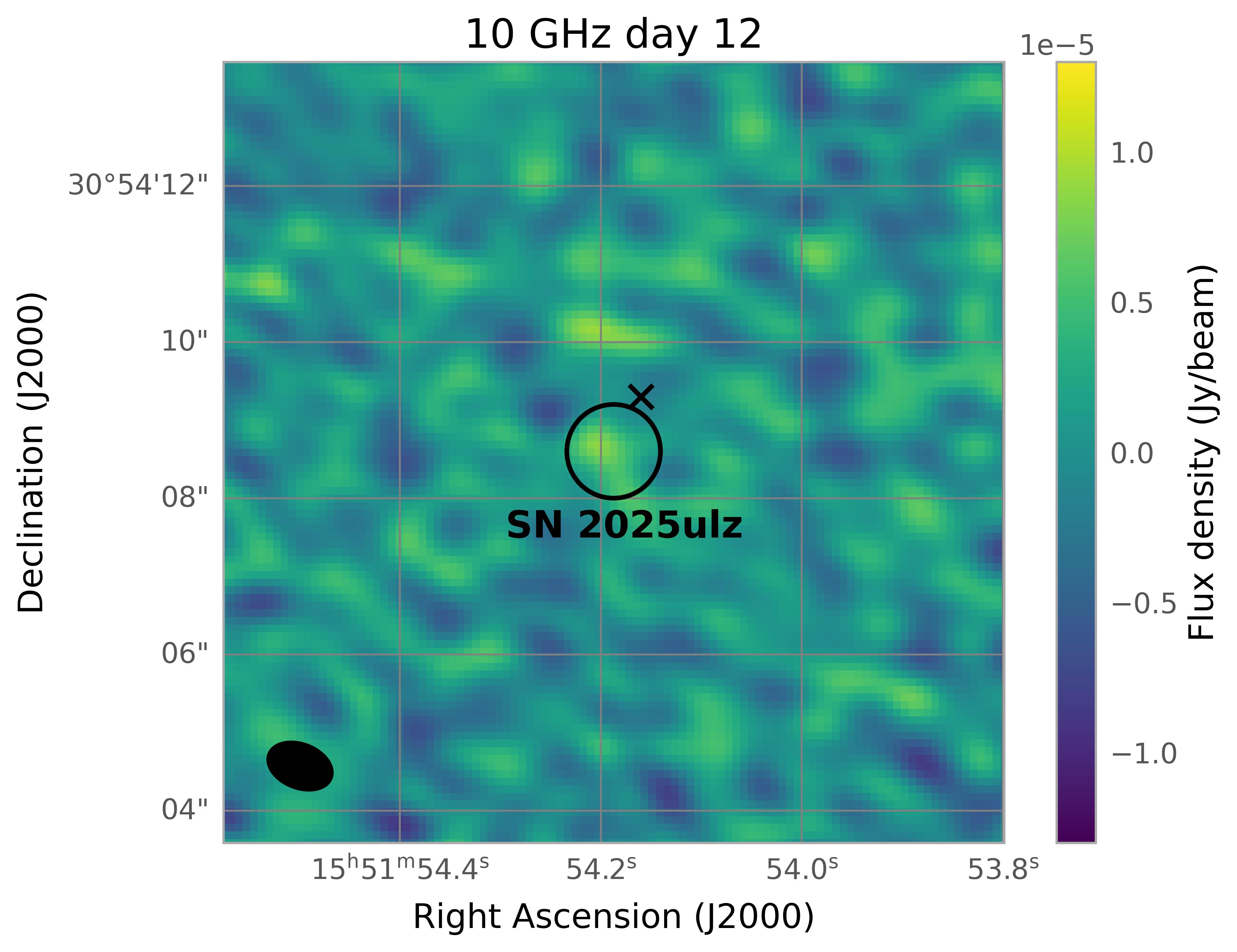}
\includegraphics[width=0.33\textwidth]{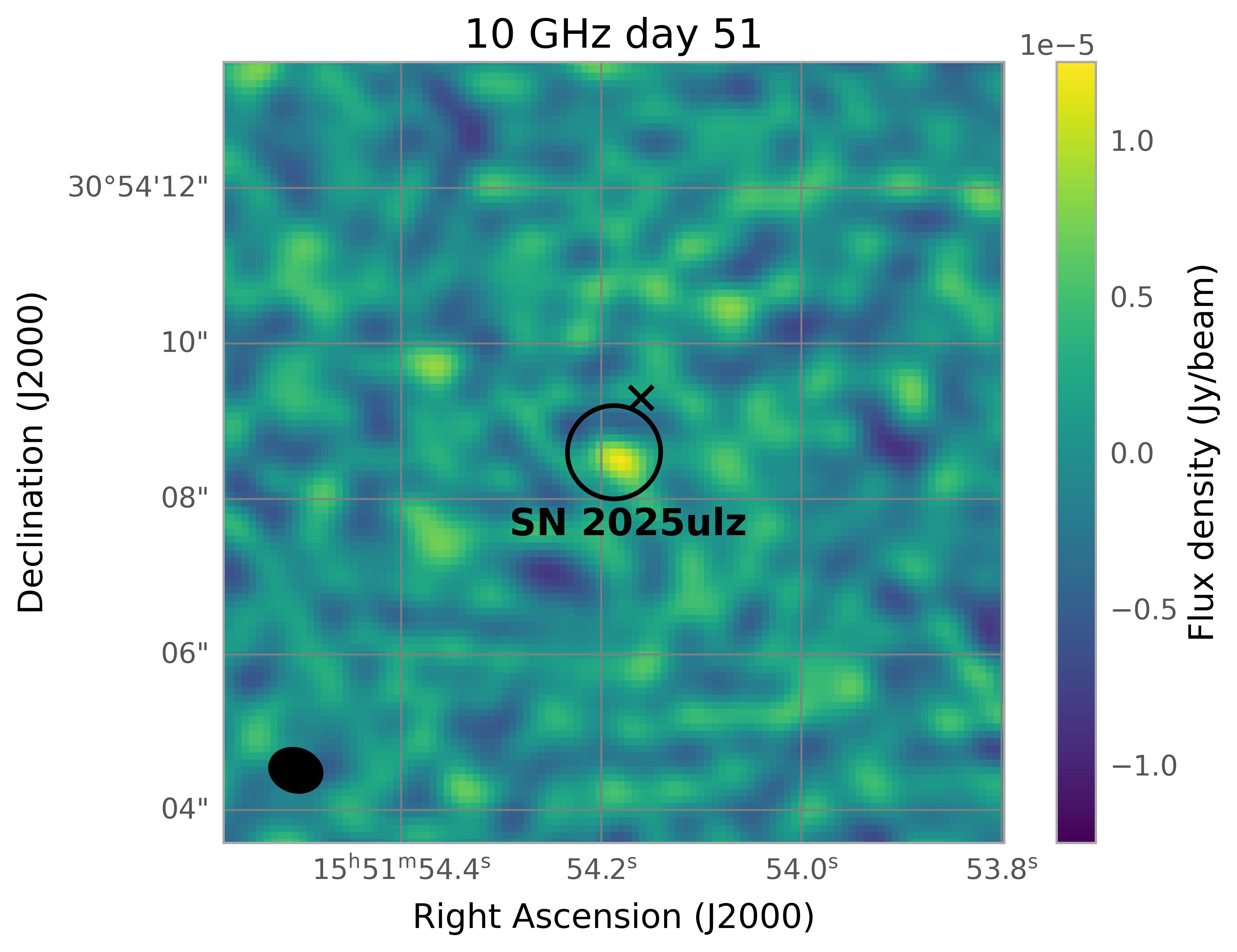}%
}
\hbox{
\includegraphics[width=0.33\textwidth]{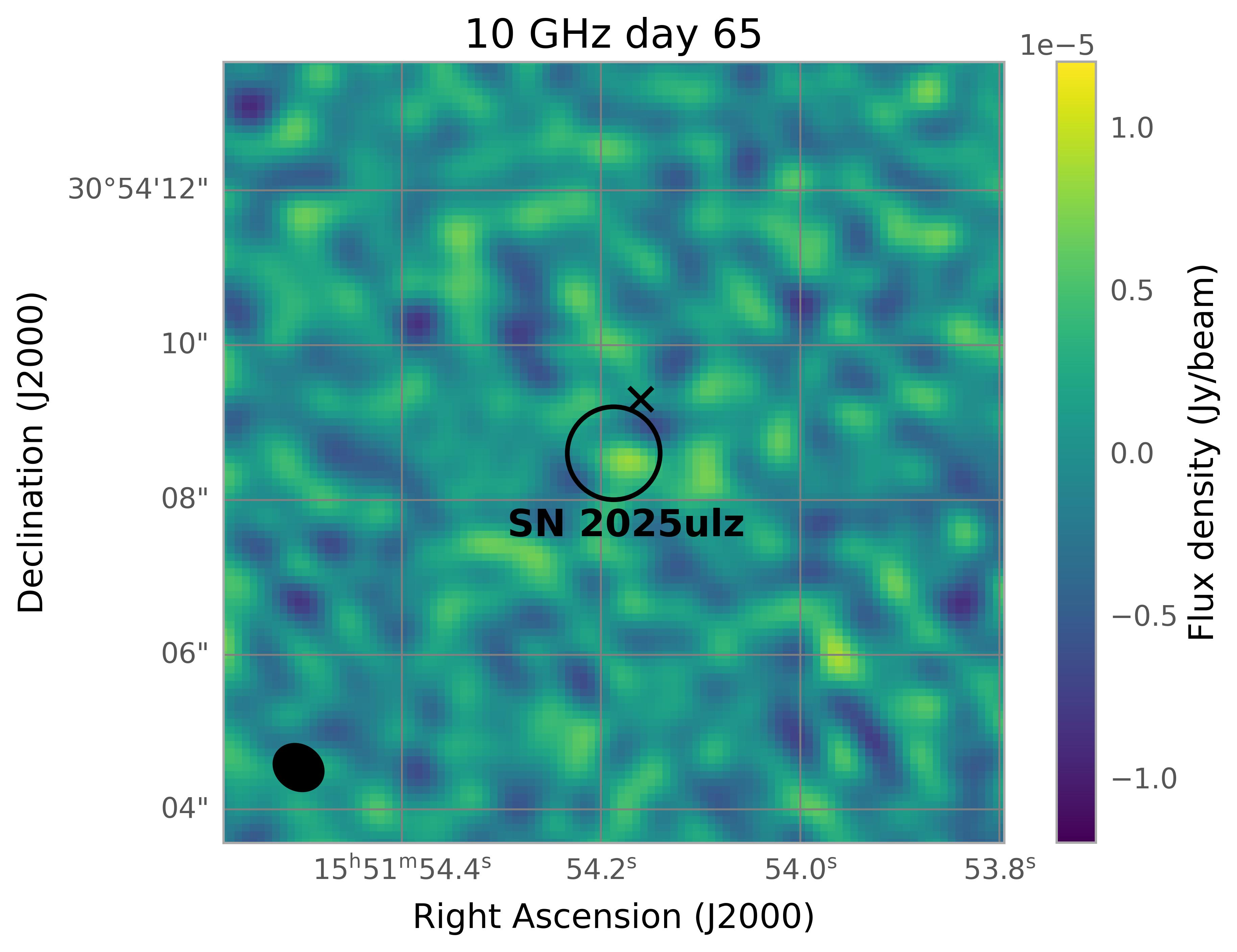}
\includegraphics[width=0.33\textwidth]{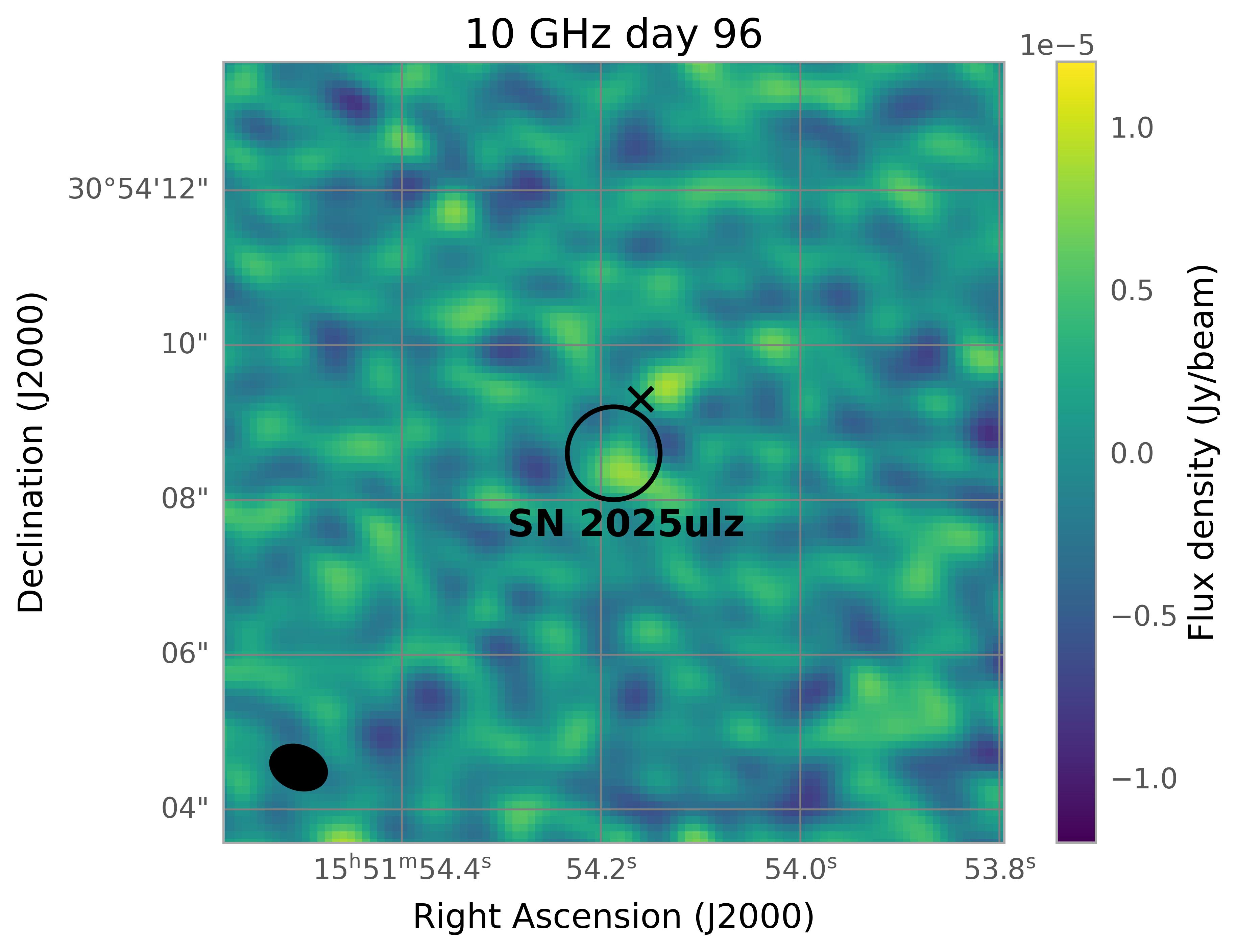}%
\includegraphics[width=0.33\textwidth]{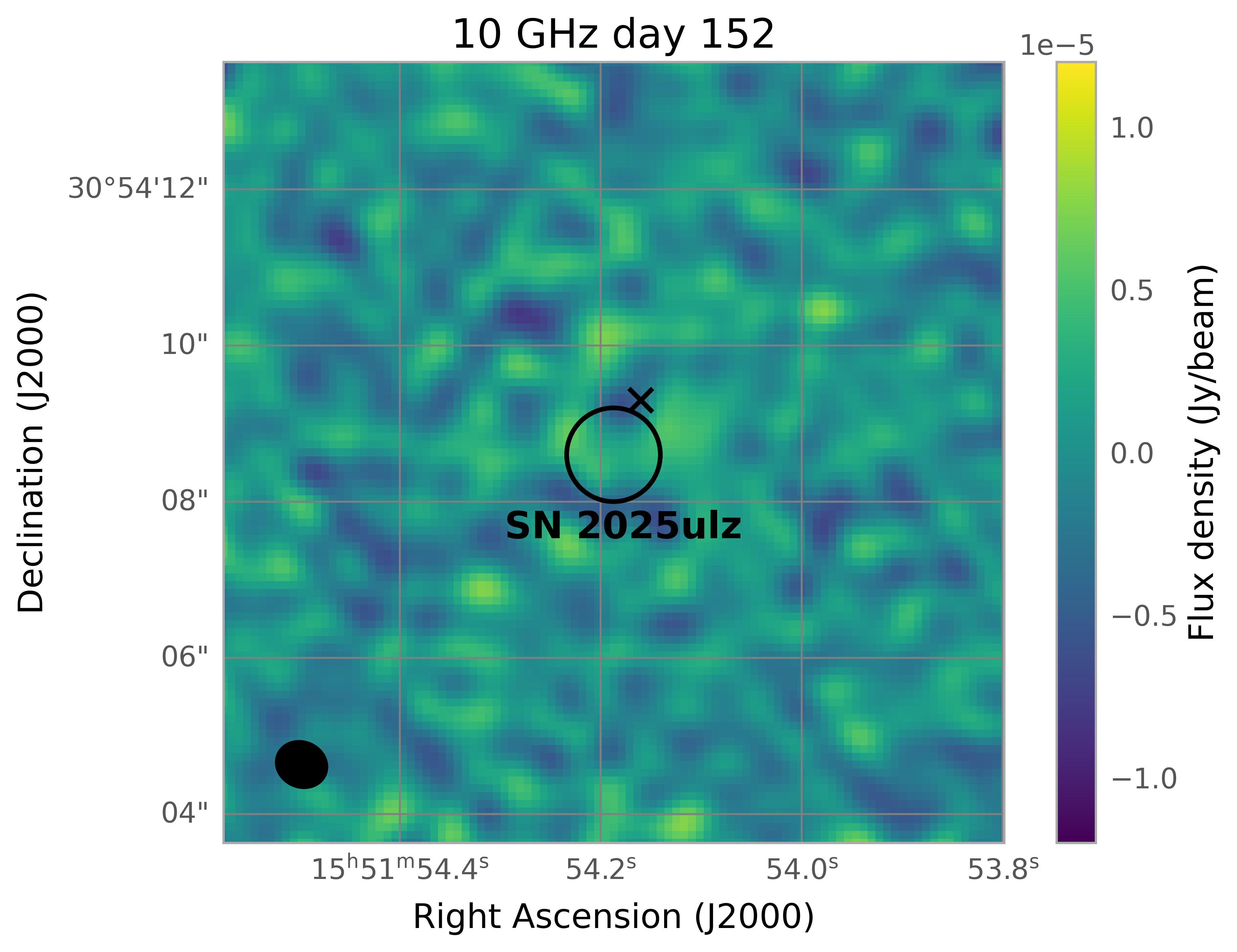}}
\caption{Observations of the SN\,2025ulz field carried out at $\nu\approx 10$\,GHz under our JAGWAR program with the VLA in its CnB (top left panel) and B (other panels) configurations. In all panels, the black circle marks a region of radius equal to the nominal FWHM of the VLA synthesized beam at 10\,GHz and in B configuration (0.60\,$\arcsec$), centered on the position of SN\,2025ulz as determined by the brightest pixel in HST images.  In Table \ref{tab:a}, we report the peak radio flux densities measured within this circular region (if above $3\times$RMS) and their positional offset and PA from the HST position of SN\,2025ulz. The actual VLA beam is plotted as a black ellipse in each panel. The optical position of the host galaxy (SDSS J155154.16+305409.3) center is marked with a black cross. Map colors span flux density values in between $\approx -5\times$\,RMS and $+5\times$\,RMS (see Table \ref{tab:a} for the RMS value of each image). Emission from a faint counterpart from SN\,2025ulz is detected at the $\approx 4.5\sigma$ level on day 51 and remains visible at the $\approx 3\sigma$ level up day 96.}
\label{fig:Xband}
\end{figure*}

\begin{figure*}
\centering
\hbox{
\includegraphics[width=0.33\textwidth]{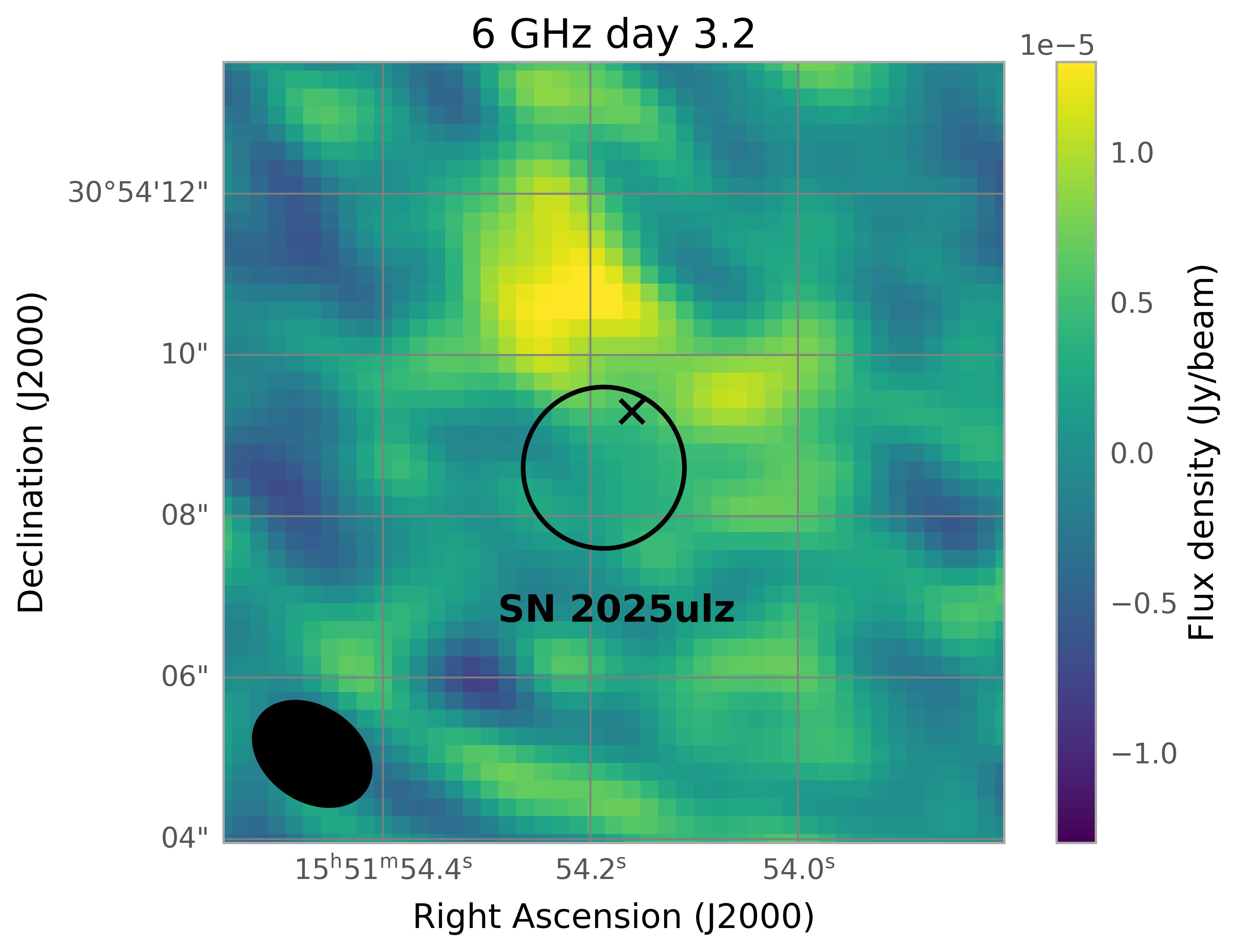}
\includegraphics[width=0.33\textwidth]{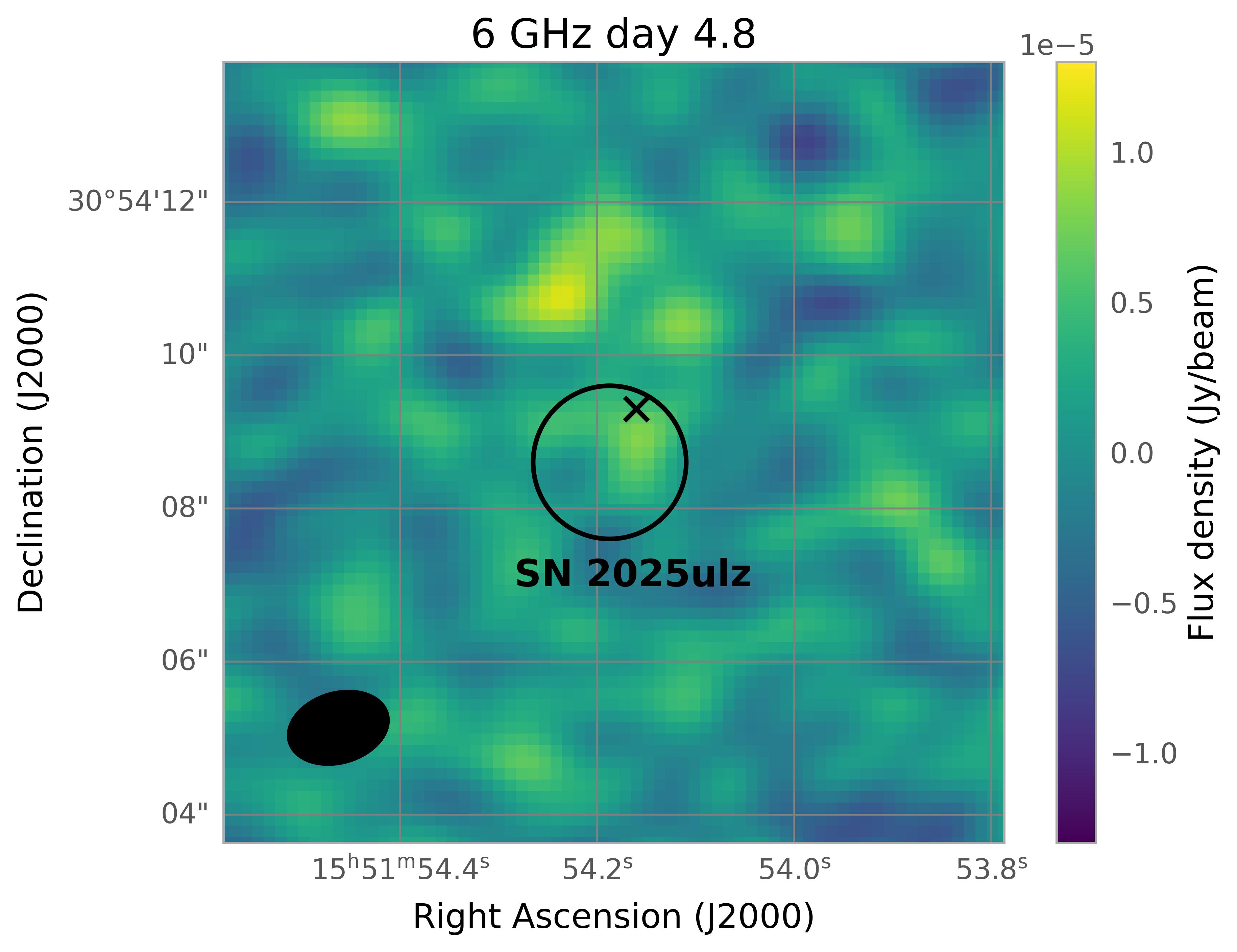}
\includegraphics[width=0.33\textwidth]{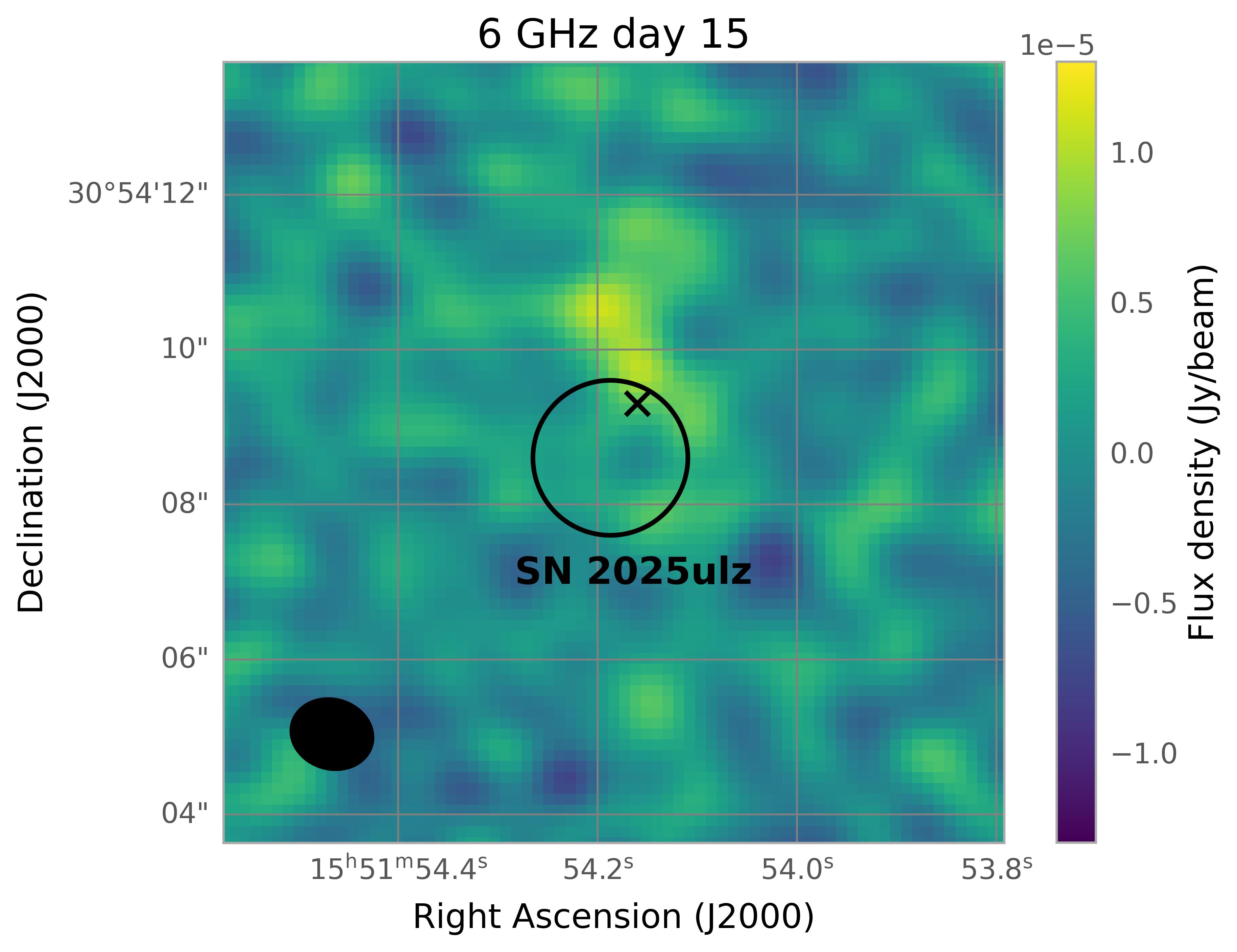}}
\hbox{
\includegraphics[width=0.33\textwidth]{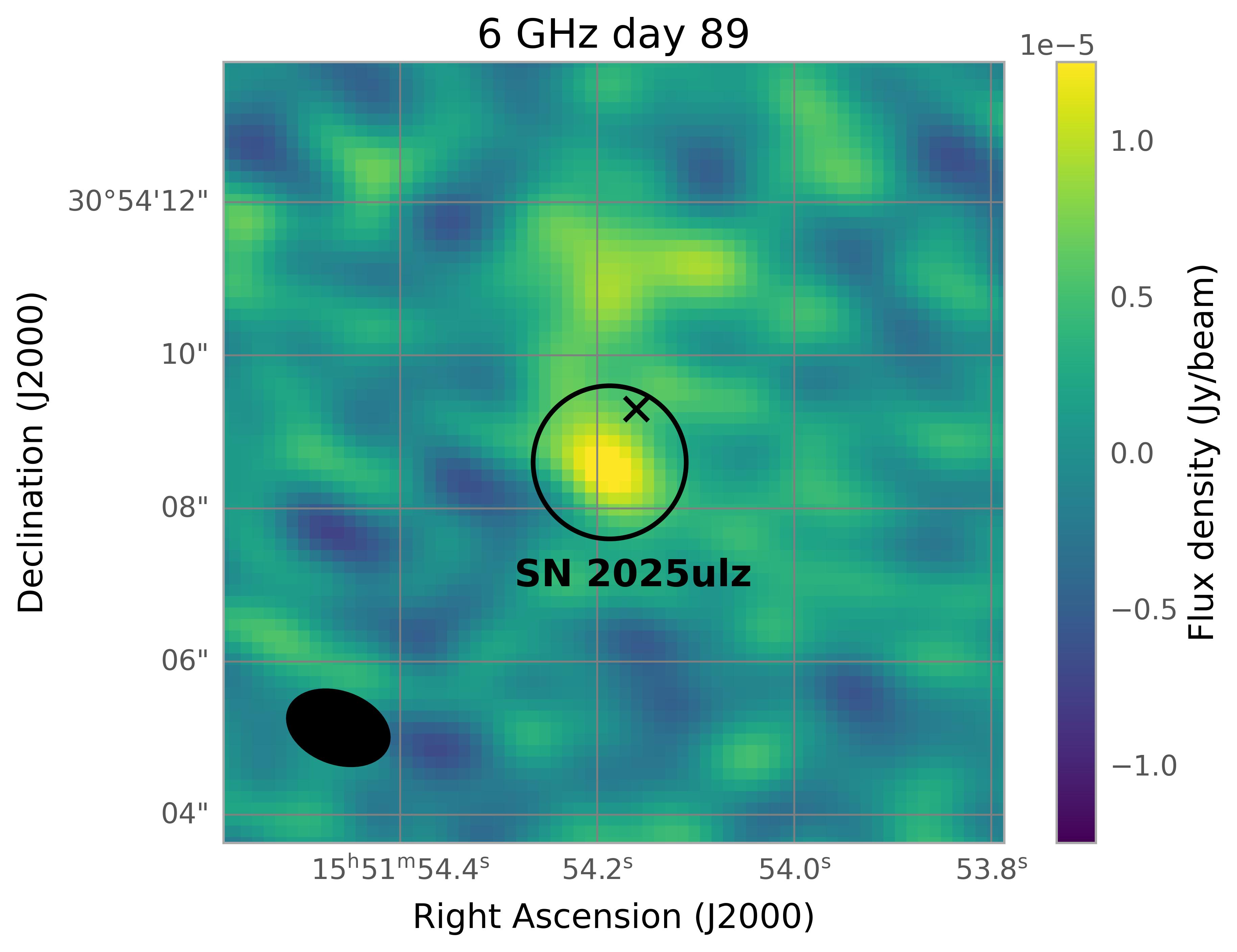}
\includegraphics[width=0.33\textwidth]{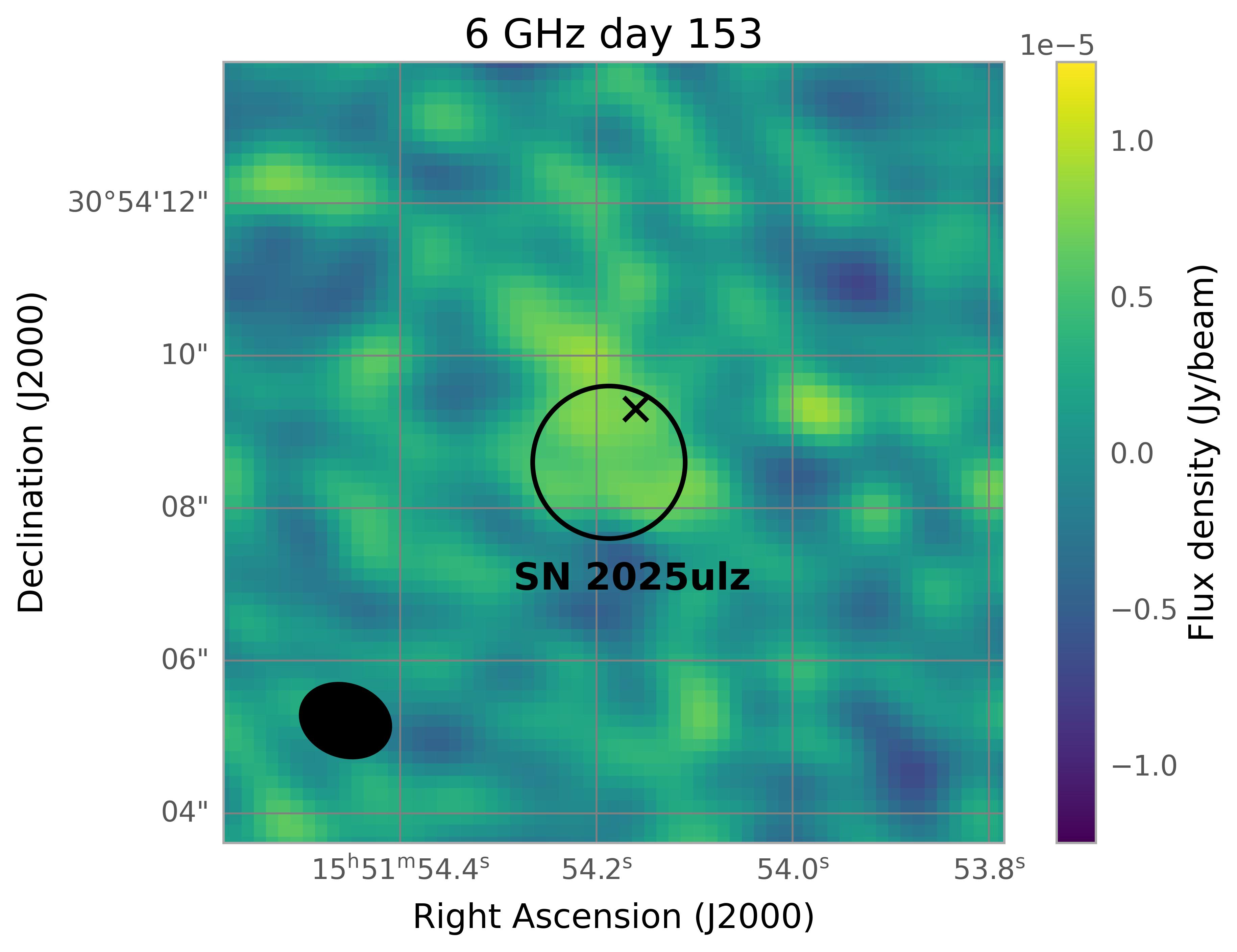}
\includegraphics[width=0.33\textwidth]{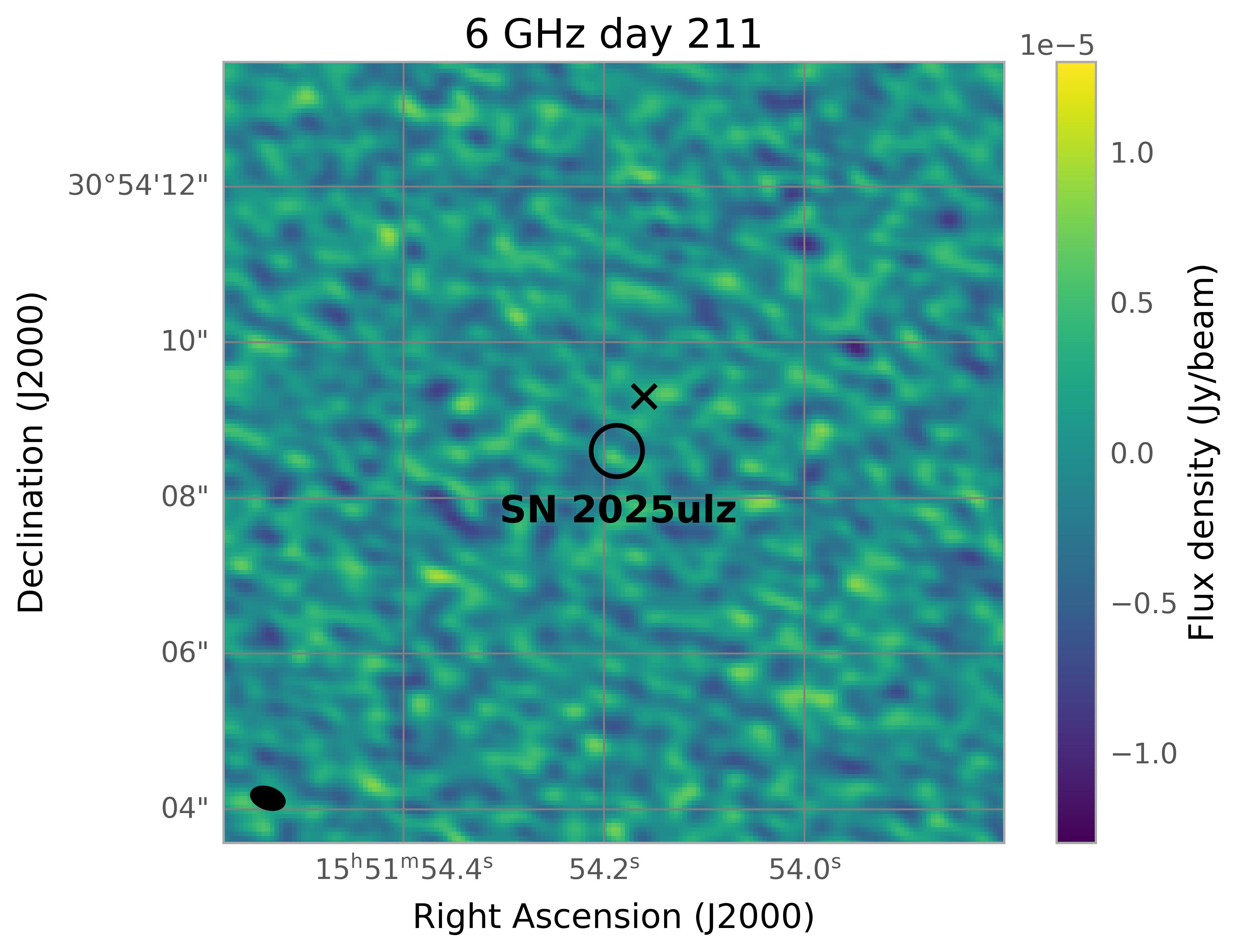}}
\caption{Observations of the SN\,2025ulz field carried out at $\nu\approx 6$\,GHz under our JAGWAR program with the VLA in its CnB (top-left and top-center panels), B (top-right, bottom-left, and center-left panels), and A (bottom-right panel) configurations. In all panels except for the bottom-right one, the black circle marks a region of radius equal to the nominal FWHM of the VLA synthesized beam at $\nu\approx 6$\,GHz and in B configuration (1.0\,$\arcsec$), centered on the position of SN\,2025ulz as determined by the brightest pixel in HST images.  This circle ha a radius of $0.33\arcsec$ (nominal VLA FWHM at $\nu\approx 6$\,GHz and in A configuration) in the bottom-right panel. In Table \ref{tab:a}, we report the peak radio flux densities measured within this circular region (if above $3\times$RMS) and their positional offset and PA from the HST position of SN\,2025ulz. The actual VLA beam is plotted as a black ellipse in each panel. The optical position of the host galaxy (SDSS J155154.16+305409.3) center is marked with a black cross. Map colors span flux density values in between $\approx -5\times$\,RMS and $+5\times$\,RMS (see Table \ref{tab:a} for the RMS value of each image). Emission from a faint counterpart from SN\,2025ulz is detected at the $\approx 5\sigma$ level on day 89. }
\label{fig:Cband}
\end{figure*}

\begin{figure*}
\centering
\hbox{
\includegraphics[width=0.25\textwidth]{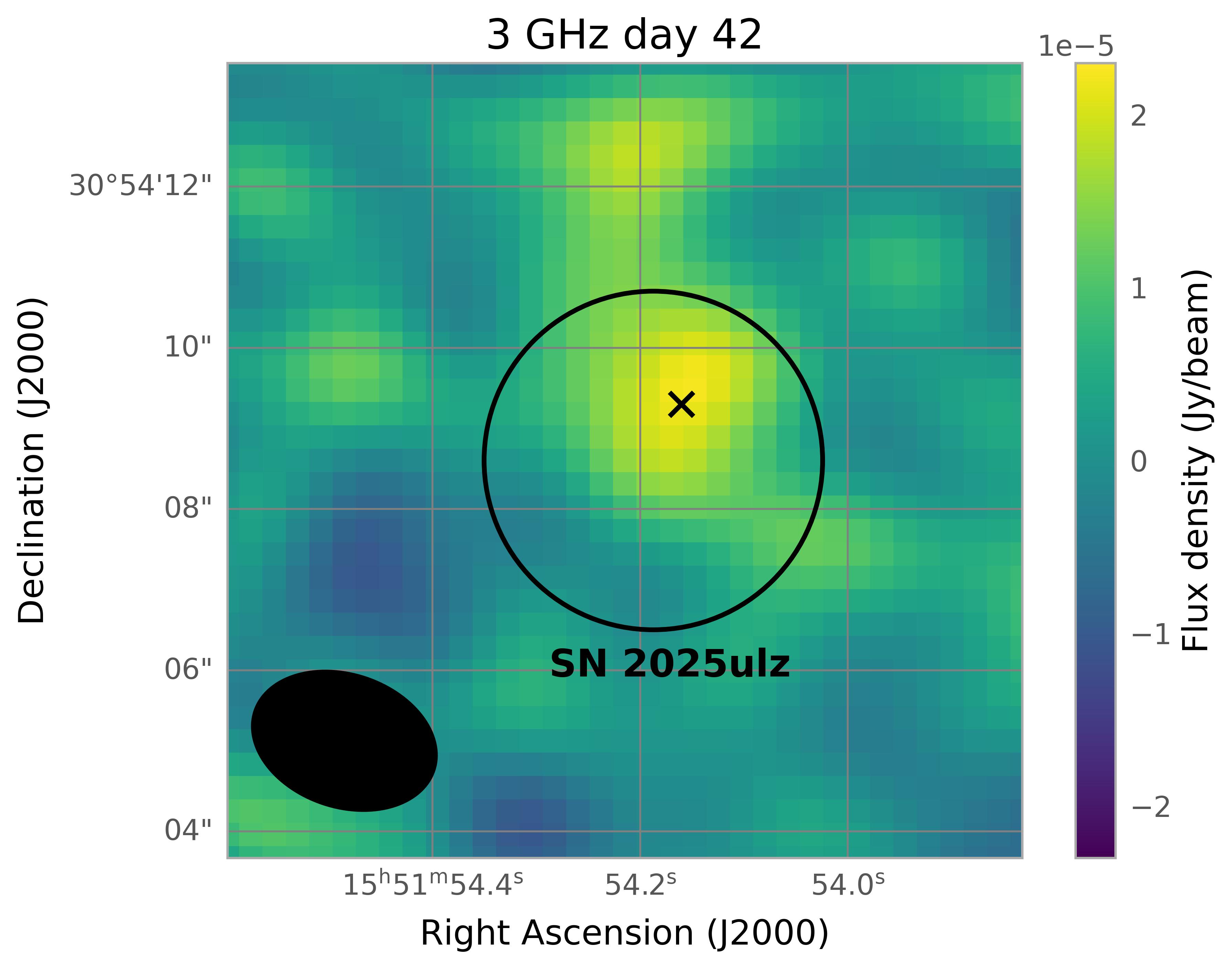}%
\includegraphics[width=0.25\textwidth]{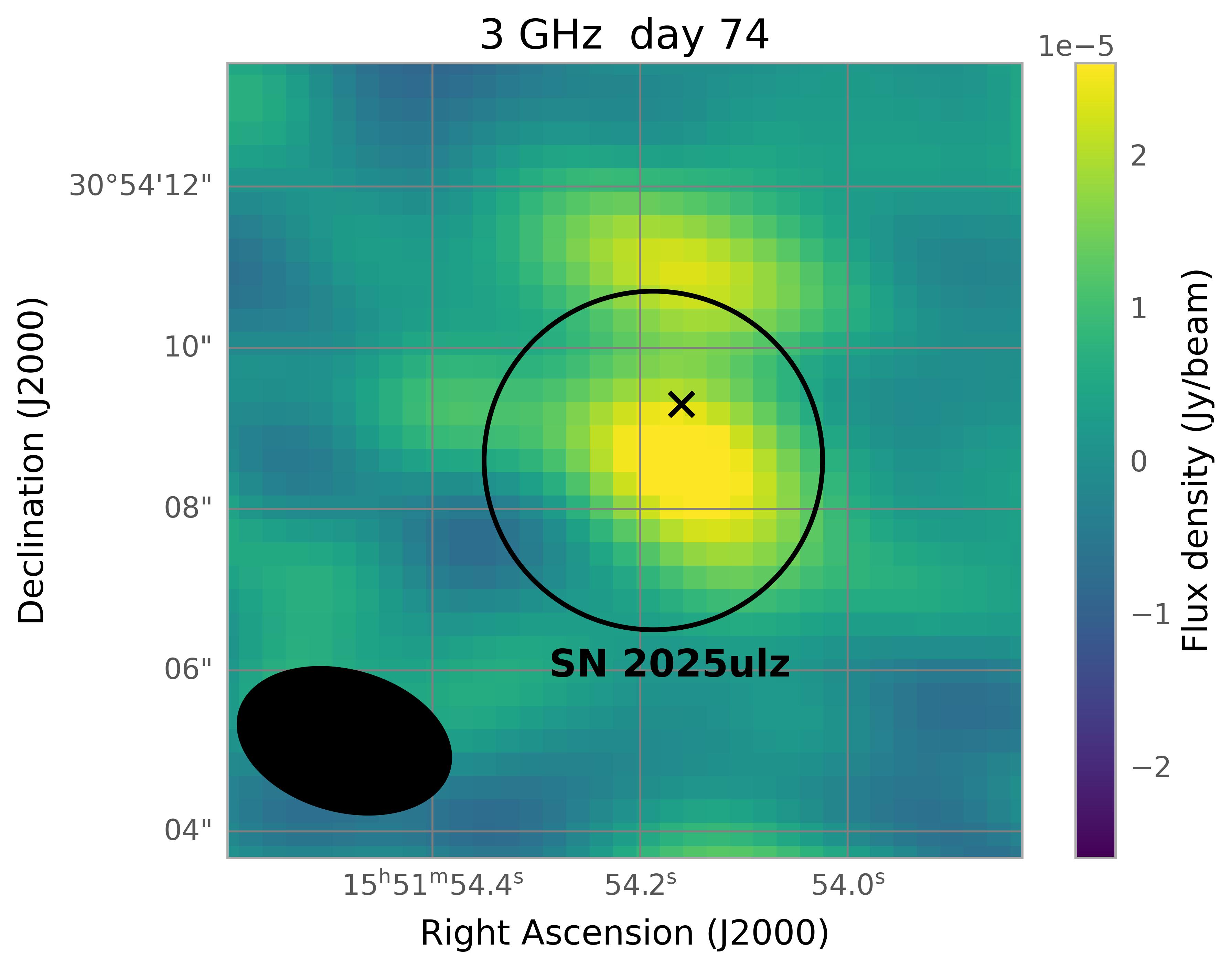}
\includegraphics[width=0.25\textwidth]{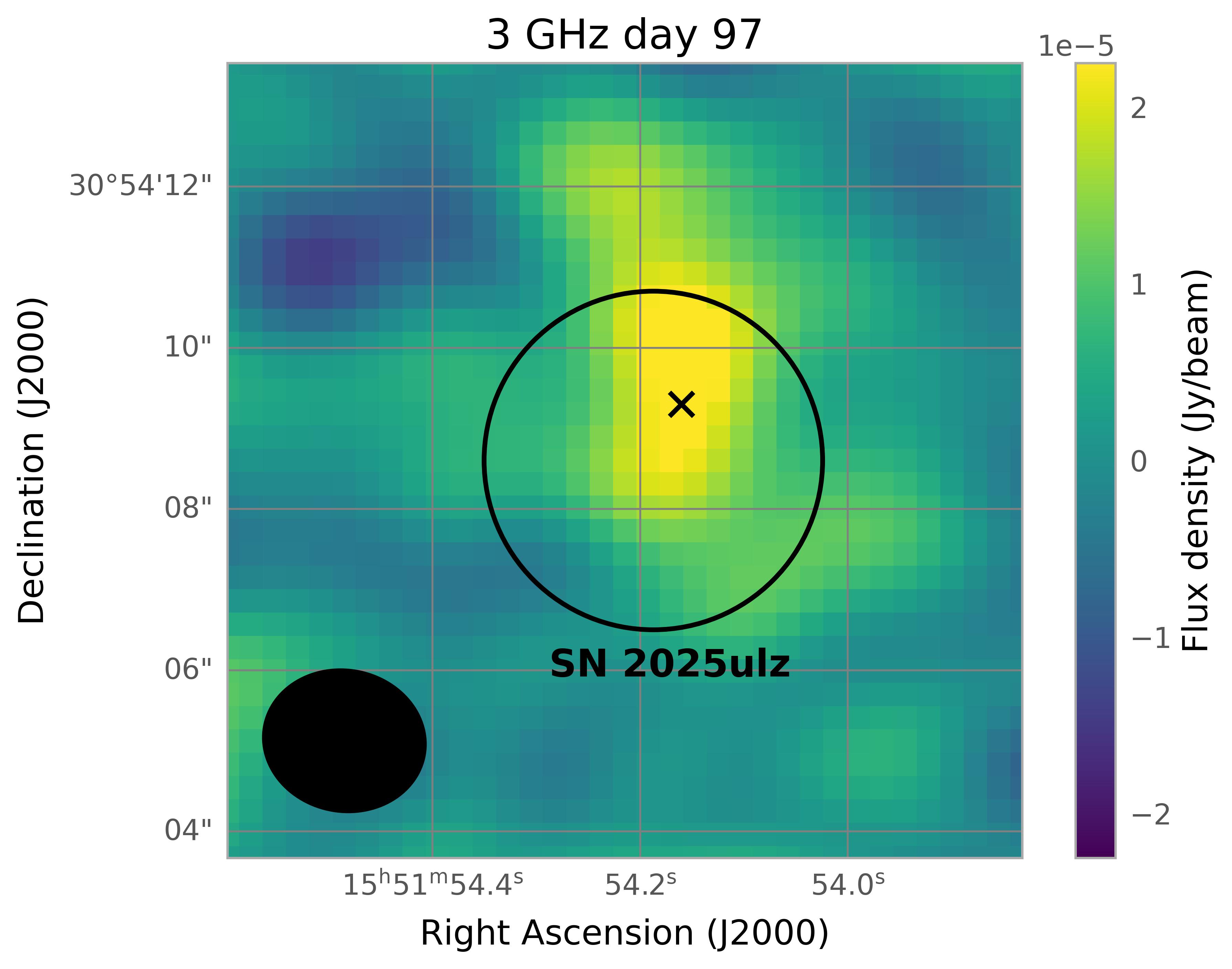}
\includegraphics[width=0.25\textwidth]{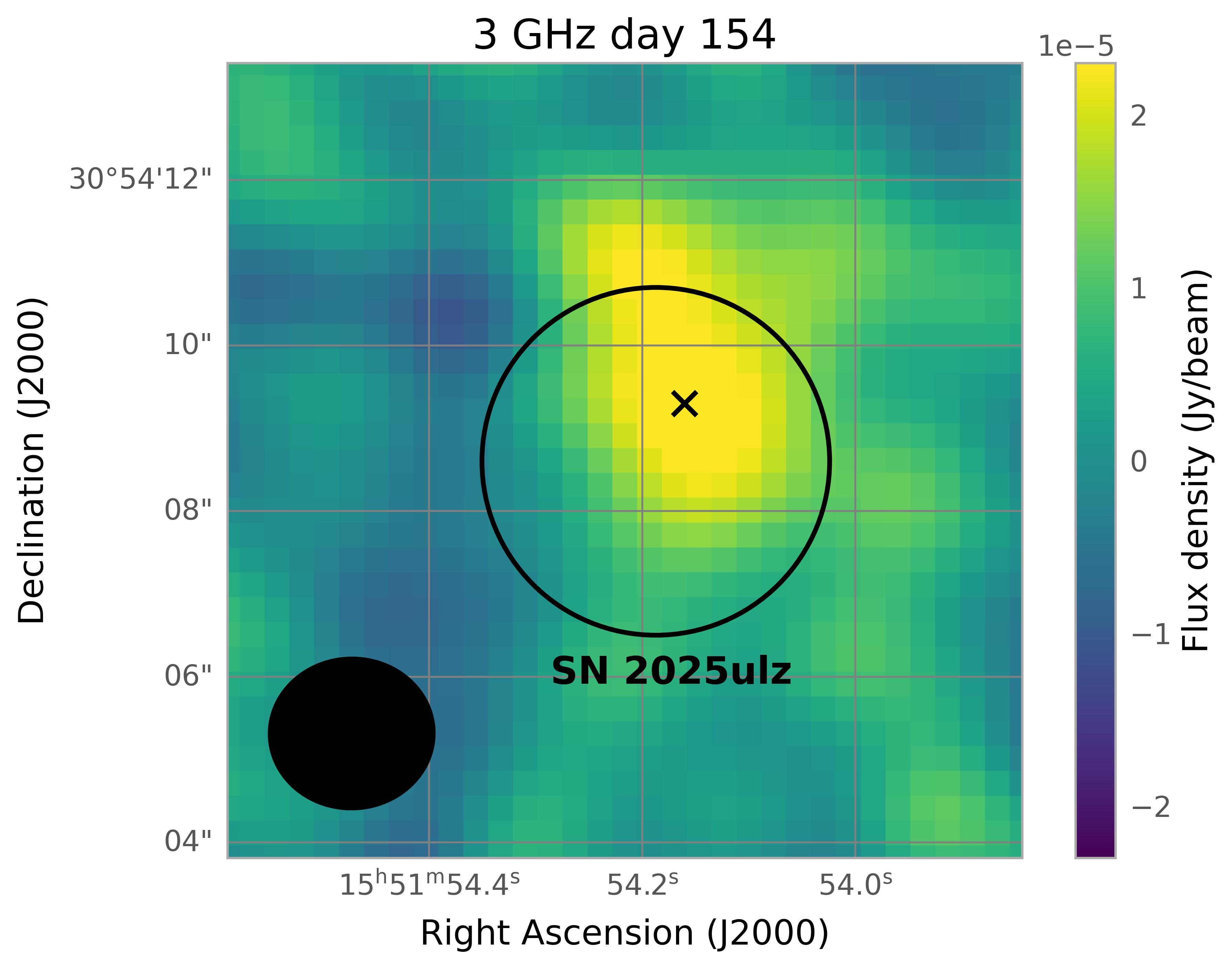}}
\caption{Observations of the SN\,2025ulz field carried out at $\nu\approx 3$\,GHz under our JAGWAR program with the VLA in its B configuration. In all panels, the black circle marks a region of radius equal to the nominal FWHM of the VLA synthesized beam at $\nu\approx 3$\,GHz and in B configuration  (2.1$\arcsec$), centered on the position of SN\,2025ulz as determined by the brightest pixel in HST images.  In Table \ref{tab:a}, we report the peak radio flux densities measured within this circular region (if above $3\times$RMS) and their positional offset and PA from the HST position of SN\,2025ulz. The actual VLA beam is plotted as a black ellipse in each panel. The optical position of the host galaxy (SDSS J155154.16+305409.3) center is marked with a black cross. Map colors span flux density values in between $\approx -5\times$\,RMS and $+5\times$\,RMS (see Table \ref{tab:a} for the RMS value of each image). While at all epochs the observed flux at the position of SN\,2025ulz is dominated by host galaxy emission, there is a hint for a potential brightening due to the contribution of the AT\,2025ulx radio counterpart at day 74.}
\label{fig:Sband}
\end{figure*}

\end{document}